\documentclass[aps,twocolumn,superscriptaddress,showpacs,nofootinbib]{revtex4-2}
\usepackage{graphicx,amsmath,amssymb,color}

\newcommand{\boldr}{{\bf{r}}}
\newcommand{\boldv}{{\bf{v}}}
\newcommand{\boldj}{{\bf{j}}}
\newcommand{\boldS}{{\bf{S}}}

\newcommand{\zhat}{{\mbox{\boldmath ${\hat z}$}}}

\def\blue#1{\textcolor{blue}{#1}}

\begin{document}

\title{Superfluid drain vortex}

\author{Wandrille Ruffenach}
\affiliation{D\'epartement de Physique, \'Ecole Normale Sup\'erieure de Lyon,
46 All\'ee d’Italie, F 69342 Lyon Cedex 07, France}
%\email{wandrille.ruffenach@ens-lyon.fr}

\author{Luca Galantucci}
\affiliation{School of Mathematics, Statistics and Physics, 
Newcastle University, Newcastle upon Tyne, 
NE1 7RU,  United Kingdom}
\affiliation{Istituto per le Applicazioni del Calcolo `M. Picone', (IAC-CNR),
via dei Taurini, 19, 00185 Roma, Italy}
\author{Carlo F. Barenghi}
%\email{carlo.barenghi@newcastle.ac.uk}
\affiliation{School of Mathematics, Statistics and Physics, 
Newcastle University, Newcastle upon Tyne, 
NE1 7RU,  United Kingdom}

\begin{abstract}
Drain vortices are among the most common vortices observed
in everyday life, yet their physics is complex due to the competition of
vorticity's transport and diffusion, and the presence of viscous layers and a
free surface.
Recently, it has become possible to study experimentally
drain vortices in superfluid liquid helium, a fluid in which
the physics is simplified by the absence of viscosity and the quantisation 
of the circulation. Using the Gross-Pitaevskii equation, we make a
simple model of the problem which captures
 the essential physics ingredients,
showing that the superfluid drain vortex consists 
of a bundle of vortex lines
which twist, thus strengthening the axial flow into the drain.
\end{abstract}

%\keywords{vortex, superfluid, helium, drain, suction}

\maketitle

\section{\label{sec:introduction}Introduction}
%\label{sec:introduction}

Quantised vorticity is a distinguishing property of superfluids,
indeed hundreds of papers have been written on this subject
since Vinen's detection of single quanta of circulation in
superfluid helium \cite{Vinen1961}.
It is therefore remarkable that, until recently,
very little attention has been dedicated in the superfluid context
to {\it drain vortices}
(also called {\it suction vortices} or {\it bathtub
vortices}).
Combining azimuthal motion around an axis
with radial/axial inflow into a hole, drain vortices
are familiar to everybody because they can
be easily created in a kitchen or bathroom sink filled with water.
Familiarity is not the same as physical understanding, however: 
boundaries and a free surface create subtle
Ekman layers, drainpipe axial flows and upwellings
which, even for water drain vortices, have been studied 
only recently \cite{Andersen,Bohling}.

We consider liquid helium ($^4$He) 
at temperature $T < T_{\lambda}$ where
$T_{\lambda}=2.17~\rm K$ is the critical temperature at saturated
vapour pressure.  In this low temperature
regime, liquid helium consists of two inter-penetrating
fluid components, an inviscid
superfluid and a viscous normal fluid, whose proportions 
strongly depend on $T$.
The normal fluid fraction tends to one for $T \to T_{\lambda}$ 
and to zero for $T \to 0$; viceversa, the superfluid fraction tends to zero
for $T \to T_{\lambda}$ and to one for $T \to 0$
(in practice, at $T=1~\rm K$ the superfluid fraction is already more than 
$99 \%$).

As mentioned, the key property of the superfluid component is
that any vorticity is concentrated in thin vortex lines of fixed
circulation $\kappa=h/m$ where $h$ is Planck's constant and $m$ is the mass
of one helium atom. Therefore we expect
that, in the configuration of a drain vortex,
the flow pattern of the normal fluid should be (in the first approximation)
similar to that of water, including viscous layers and
a continuous vorticity field which fills the system, whereas the
vorticity of the superfluid should be confined to a central cluster of
vortex lines.  However, since the vortex lines scatter the thermal 
excitations (phonons and rotons) which make up the normal fluid, 
there should also be a mutual friction force 
between normal fluid and superfluid components, whose precise effects 
are difficult to guess (the friction depends on the density of the vortex
lines and their velocity difference with respect to the normal fluid).
We also expect that the flow
pattern should depend on whether the drain vortex is created mechanically 
(e.g. using a propeller) or thermally (e.g. using a heater), as the
induced superfluid and normal fluid velocities will be
parallel or antiparallel respectively. Finally, unless the container is very
large, details of any fluid reinjection into the system will be important.
In summary, the superfluid drain vortex problem contains many physical 
ingredients which may combine in a nontrivial way, even before considering 
the presence of a free surface, 
which may deepen creating a funnelling drainpipe.

Experimentally, the problem has been tackled recently
by Yano and collaborators \cite{Yano,Matsumura,Obara}. 
Using a rotor, they created a drain vortex in helium 
at $T=1.6~\rm K$ (corresponding to a superfluid fraction of $83\%$). 
By measuring the attenuation of second sound,
they showed that the drain vortex consists of a cluster of
approximately $10^4$ quantised vortex lines which accumulate in a
narrow central region near the axis of symmetry above the drain hole.
The experiment was followed up by numerical simulations \cite{Inui}
based on the Vortex Filament Model (VFM). These simulations
determined the evolution of seeding vortex lines at nonzero
temperatures in the presence of a prescribed normal fluid in the
shape of a Rankine vortex with a superimposed constant axial flow into
the drain hole.

In this work, we consider the superfluid drain vortex
problem in its simplest form: a vortex cluster in the
presence of a drain flow in a pure superfluid at $T=0$, in the absence of
a free surface.
We model the problem using the Gross-Pitaevskii Equation (GPE) for a 
weakly-interacting gas of bosons \cite{Primer,StringariPitaevskii}.
The GPE is idealized for helium (which is a liquid, not a dilute gas), 
and, since its numerical solution requires the
resolution of length scales smaller than the vortex core, our calculation
is necessarily limited to small-scale vortex configurations.
Still, our model captures the essential physical ingredients, namely
the dynamics of vortex lines in a nontrivial three-dimensional
geometry which includes
a draining superflow and the presence of boundaries, 
ingredients which were not accounted in the VFM approach \cite{Inui}. Our work
is articulated as follows:
Section~(\ref{sec:model}) describes our model and the numerical methods employed
and Section~(\ref{sec:results}) presents our results.

\section{\label{sec:model}Model}
%\label{sec:model}

\subsection{\label{sec:governing}Governing equations}
%\label{sec:governing}

It is convenient to write the GPE in dimensionless form. We use
the healing length $\xi=\hbar/\sqrt{g m n_0}$ and the speed of sound
$c=\sqrt{n_0 g/m}$ as units of length and speed respectively, $\tau=\xi/c$
as unit of time, the density of a uniform condensate, $n_0$, as
the unit of density, and express the trapping potential in units of 
the chemical potential $g n_0$, $g$ being the interaction parameter,
$m$ the mass of one atom, and $\hbar=h/(2 \pi)$ the reduced 
Planck's constant.  The resulting dimensionless GPE is

\begin{equation}
i \frac{\partial \psi}{\partial t}=
-\frac{1}{2} \nabla^2 \psi + \vert \psi \vert^2 \psi +V \psi,
\label{eq:GPE}
\end{equation}

\noindent
where $\psi(\boldr,t)$, $n(\boldr,t)=\vert \psi(\boldr,t)\vert^2$,
$V(\boldr)$, $\boldr$
and $t$ are the dimensionless wavefunction, density,
trapping potential, position and time respectively (hereafter all
quantities are meant to be dimensionless).
During the evolution, Eq.~(\ref{eq:GPE}) conserves the total
number of atoms, $N$, in the volume $\cal V$ of the system,
and the energy, $E$, given respectively by

\begin{equation}
N=\int_{\cal V} n \, d^3\boldr,
\label{eq:N}
\end{equation}

\begin{equation}
E=\int_{\cal V} \left( \frac{1}{2} \vert \nabla \psi \vert^2 +
\frac{1}{2} \vert \psi \vert^4 + V \vert \psi\vert^2 \right) d^3\boldr.
\label{eq:E}
\end{equation}
We solve the three-dimensional
GPE numerically in the computational domain
$x_{min} \le x \le x_{max}$, $y_{min} \le y \le y_{max}$,
$z_{min} \le z \le z_{max}$
%, where $x_{min}=y_{min}=-35$,
%$x_{max}=y_{max}=35$, $z_{min}=-30$ and $z_{max}=30$,
using the $4^{\rm th}$ order Runge-Kutta method in time and
centred differences in space; the spatial discretization consists of
$N_x$, $N_y$ and $N_z$ discretization points in the $x$, $y$ and
$z$ directions respectively.
%$N_x=N_y=140$ points in the $x$ and $y$ directions and
%$N_z=120$ in the $z$ direction. 
%We impose the boundary condition $\psi=0$
%on the boundary of the domain.
Convergence is checked by monitoring the conservation laws; for 
typical time evolutions described in the next section, $N$ and $E$
are conserved with relative errors
$(N-N_0)/N_0 \approx 10^{-5} \, \%$ and $(E-E_0)/E_0 \approx 0.05 \, \%$
where $N_0$ and $E_0$ are the initial values.

\subsection{\label{sec:geometry}Geometry of the system}
%\label{sec:geometry}

\begin{figure}[!ht]
\centering
\includegraphics[angle=0,width=0.4\textwidth]{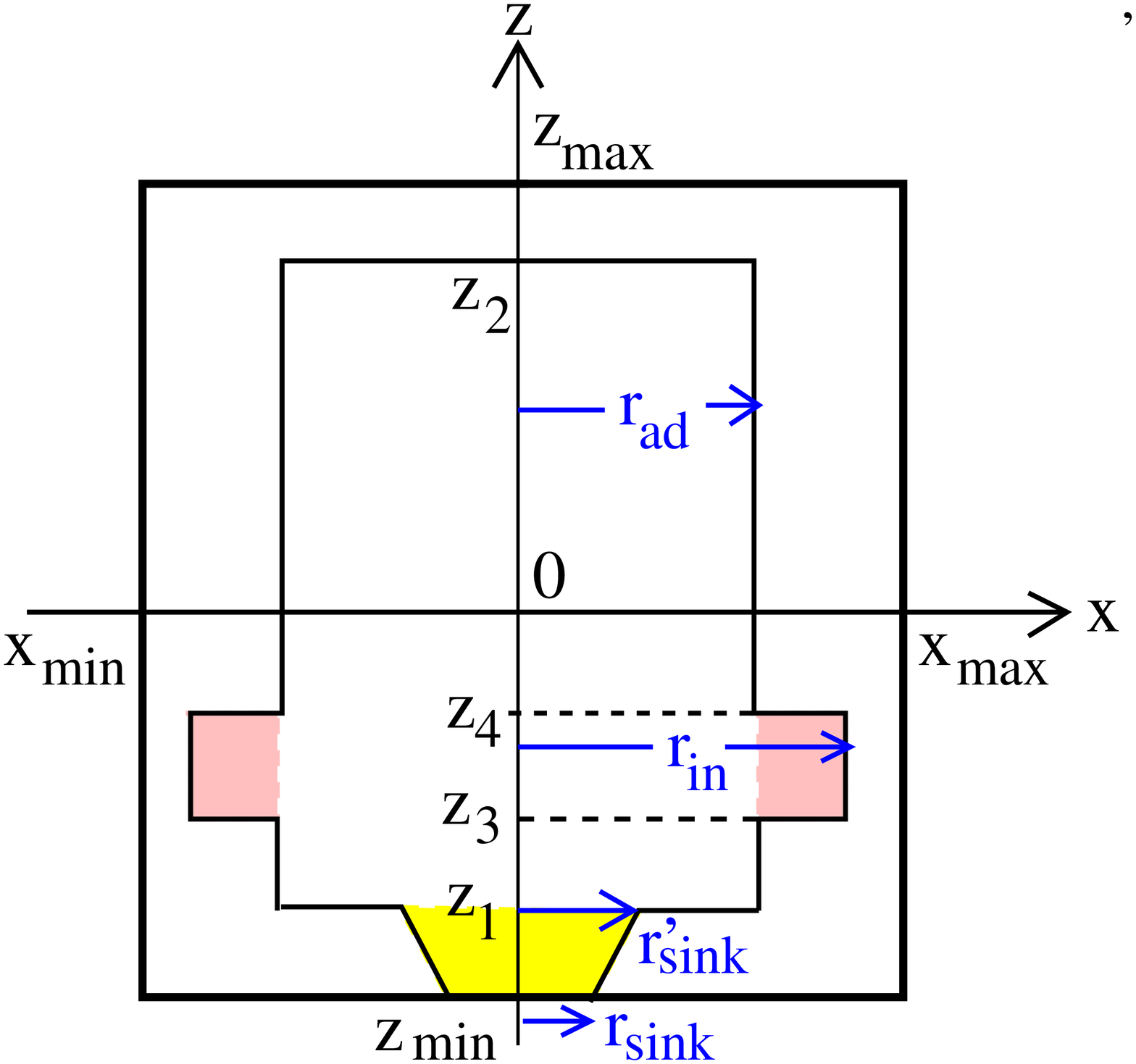}
\includegraphics[angle=0,width=0.4\textwidth]{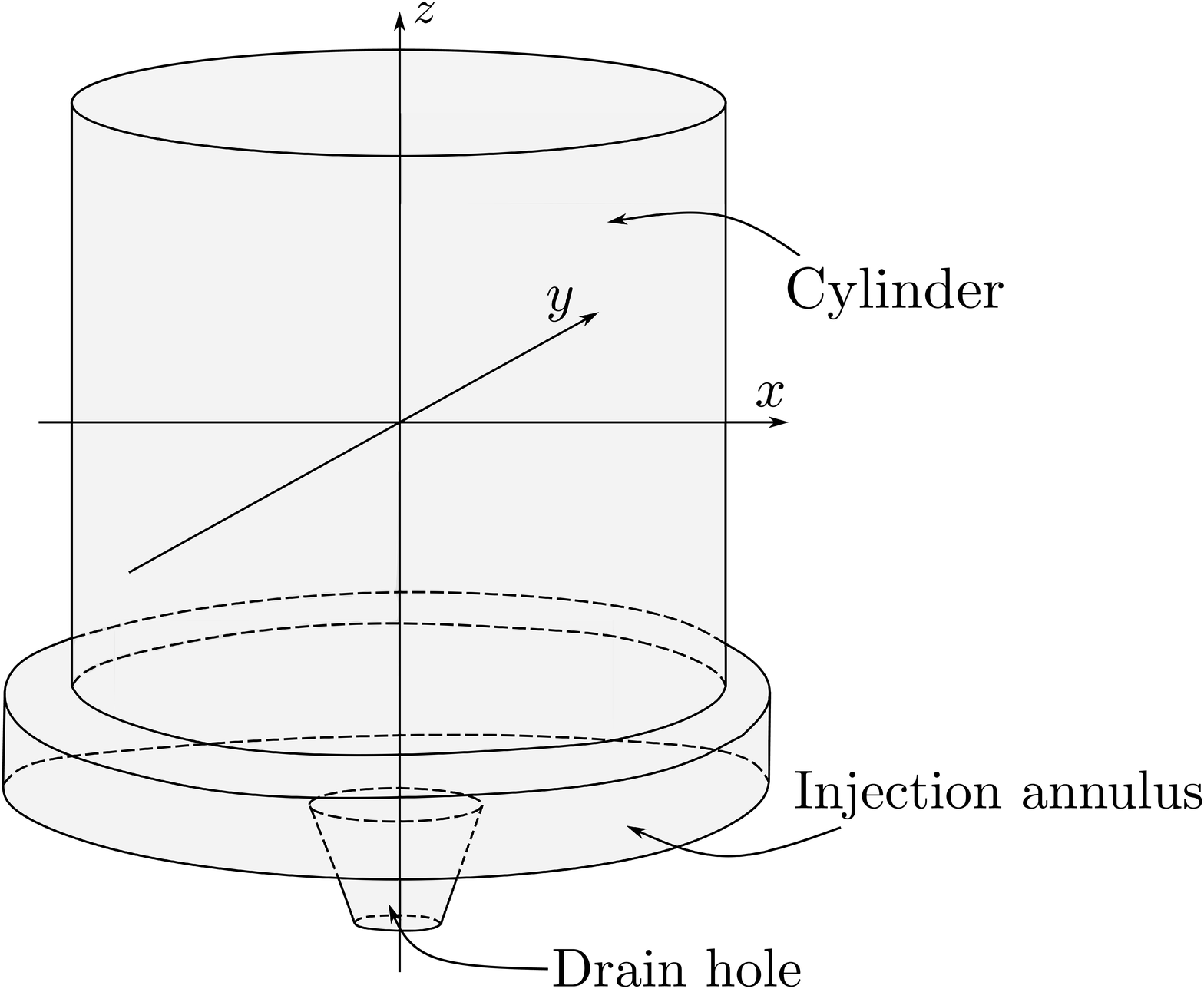}
\caption{The geometry of our numerical experiments.
Top: schematic cross section on the $y=0$ plane. The yellow and pink regions
are the drain hole and the injection annulus, respectively.
Bottom: schematic three-dimensional view for $z_3=z_1$.
}
\label{fig:1}
\end{figure}

For a system whose geometry is described in Fig.~\ref{fig:1}, 
we set $V(\boldr)=V_{trap}(\boldr)$
in Eq.~(\ref{eq:GPE}),
where the trapping potential
$V_{trap}(\boldr)$ is equal to zero inside the following three regions:
%shown in Fig.~(\ref{fig:1}):
(i) a cylinder of radius $r_{ad}$ extending from
$z_1$ to $z_2$ which represents the experimental cell;
(ii) a drain hole in the shape of
a truncated cone with top radius $r'_{sink}$ and bottom
radius $r_{sink}$ extending from $z_{min}$ to $z_1$;
(iii) an injection annulus of inner
radius $r_{ad}$ and outer radius
$r_{in}$ extending from $z_3$ to $z_4$.
Outside these three regions, we set $V_{trap}(\boldr)=10$ 
(in units of the chemical potential). This value corresponds to a high potential
barrier imposing $\psi=0$ outside the regions (i) - (iii). 
In the experiment of Yano and collaborators \cite{Yano}
the injection annulus is at the bottom of the experimental cell, 
so here we present results
for $z_3=z_1$ (i.e. the injection annulus is at the bottom of the cylinder).

\subsection{\label{sec:drain}Drain and injection}
%\label{sec:drain}

To model the drain hole, we add a negative imaginary part to the potential
$V(\boldr)$
in region (ii) (the truncated cone below the cylinder),
setting

\begin{equation}
V(\boldr)=V_{trap}(\boldr)-iV_{sink}(\boldr) \, ,
\label{eq:trap}
\end{equation}

\noindent
where $V_{sink}(\boldr)$ is equal
to a positive constant $V_0$ in the drain hole
and zero elsewhere.
Between $t$ and $t+\Delta t$, this negative imaginary potential
removes $\Delta N(t)$ atoms from the drain hole, depleting the density
in that region. From Eq.~(\ref{eq:GPE}) it is in fact possible to deduce 
the following continuity equation 
via the Madelung transformation \cite{Primer}:
\begin{equation}
\frac{\partial n}{\partial t} + \nabla \cdot \left ( n \mathbf{v}\right ) = -2 n V_{sink}\, ,
\label{eq:continuity_imag_V}
\end{equation}
where $\mathbf{v}=\mathbf{j}/n$ is the velocity of particles and 
the current $\mathbf{j}$ is defined as follows:
\begin{equation}
\boldj=n \boldv=\frac{i}{2}(\psi \nabla \psi^*-\psi^* \nabla \psi) \, .
\label{eq:current_j}
\end{equation}

By integrating Eq.~(\ref{eq:current_j}) over the whole volume, we obtain
that the rate of loss of particles into the system due to drain potential
is given by $\displaystyle \blue{-} \frac{dN}{dt} = 2 V_0 N_{sink}(t)$,
where $N_{sink}(t)$ is the number of atoms in the drain hole. 
The number of atoms $\Delta N(t)$ removed from the drain hole 
in time $\Delta t$ is hence, in the first approximation, given by
$\Delta N(t) \approx 2 V_0 N_{sink}(t) \Delta t$.
As the quantum fluid described by the GPE is barotropic 
(the pressure is proportional to the square of the density \cite{Primer}), 
the density difference arising from the atoms removal in the drain creates 
a pressure difference 
which drives a flow towards the drain (see Section~\ref{sec:results}). 
In order to conserve the total number of atoms of the system, we add
into the injection annulus the same number of atoms $\Delta N(t)$
that we have removed from
the drain hole. After time-stepping the wavefunction from
$\psi(\boldr,t)$ to $\psi(\boldr, t+\Delta t)$,
the naive approach would be to
change the density in the injection annulus 
from $n(\boldr, t+\Delta t)$ to 
$n'(\boldr,t+\Delta t) = n(\boldr, t+\Delta t)+\Delta N/{\cal V}_{in}$, where ${\cal V}_{in}$
is the volume of the injection annulus. However, we have found that
the resulting small radial discontinuity of the density at the edge 
of the injection annulus ($r=r_{ad}$) tends to destabilize the solution. 
A more stable injection is obtained if the injected density
profile is continuous. Therefore we set
$n'(\boldr,t+\Delta t) = n(\boldr, t+\Delta t) + f(r) \Delta N(t)$, where the distribution
function $f(r)$ vanishes at $r=r_{ad}$ and $r=r_{in}$, and its volume integral is normalized to one. We choose
\begin{equation}
f(r)=\frac{\sin{[\pi(r-r_{ad})/(r_{in}-r_{ad})]}}
{2 (z_4-z_3)(r^2_{in}-r^2_{ad})} \,,
\end{equation}

\noindent
where

\begin{equation}
\int_0^{2\pi} \!d\theta \int_{z_3}^{z_4} \! dz \int_{r_{ad}}^{r_{in}} \! dr \, r f(r)=1 \, .
\end{equation}

%It is instructive to physically interpret the effects of the drain hole 
%and the injection annulus using the Madelung transformation, which
%turns Eq.~(\ref{eq:GPE}) into a modified Euler equation 
%and a continuity equation \cite{Primer}.  The inclusion of drain
%and injection affects only the continuity equation, which becomes
Our particle re-injection protocol can physically be interpreted as
the inclusion of a source term $\sigma_{in}$ in the continuity equation 
Eq.~(\ref{eq:continuity_imag_V}), \textit{i.e.}

\begin{equation}
\frac{\partial n}{\partial t}+ \nabla \cdot (n \boldv)=
-2 n V_{sink}(\boldr) + \sigma_{in}(\boldr,t),
\label{eq:cont}
\end{equation}
\noindent
where the production term, $\sigma_{in}(\boldr,t)$, is nonzero only inside the
injection annulus. 
Since $V_{sink}(\boldr)$ is equal to a constant, $V_0$,
inside the drain hole and vanishes outside it, in order that
the number of atoms in the system remains the same,
$\sigma_{in}(\boldr,t)$ must satisfy

\begin{equation}
\int_{{\cal V}_{in}} \sigma_{in}(\boldr,t) \, d^3\boldr=
2N_{sink}(t) V_0 \, .
\end{equation}

%\noindent
%where $N_{sink}(t)$ is the number of atoms in the drain hole
%and ${\cal V}_{in}$ is the volume of the injection annulus.

%\subsection{\label{sec:current}Current}
%\label{sec:current}

%To describe the flow we consider the current, defined as \cite{Primer}
%\begin{equation}
%\boldj=n \boldv=\frac{i}{2}(\psi \nabla \psi^*-\psi^* \nabla \psi),
%\end{equation}

%\noindent
%where $\boldv$ is the velocity. Since the density, $n$, is close to
%unity almost everywhere in the cylinder, $\boldj$ gives us direct 
%information about the flow's velocity.
%Furthermore, knowing the current, we can define the flow rate through 
%any surface $\cal S$ as
%\begin{equation}
%Q_{\cal S}=\int_{\cal S} \boldj \cdot d\boldS.
%\end{equation}

%\subsection{\label{sec:imprinting}Vortex imprinting}
%\label{sec:imprinting}

%A vortex line is imprinted into the system in a standard way
%by multiplying the ground state wavefunction times the approximate 
%wavefunction of a vortex in a homogeneous condensate.

\section{\label{sec:results}Results}
%\label{sec:results}

\subsection{\label{sec:ground}Ground state}
%\label{sec:ground}

To find the ground state of the system, we
start from the Thomas-Fermi approximation, imposing that
$\psi_{TF}(\boldr)=\sqrt{1 -V_{trap}(\boldr)}$ if $V_{trap}(\boldr)<1$ and zero otherwise.
Without any drain flow or injection, we
integrate Eq.~(\ref{eq:GPE}) in imaginary time, replacing $t$ with $-it$
and we enforce at every time step the condition
that the total number of atoms in the system does not change.
As healing regions develop near the boundaries, the wavefunction settles
down to the desired time-independent state
which minimizes the energy.

\subsection{\label{sec:drainflow}Steady drain flow}
%\label{sec:drainflow}

We solve the GPE using the ground state as initial condition, 
imposing drain flow and injection as described in Section
\ref{sec:drain}.  After an initial
transient, we obtain a steady drain flow into the drain hole
which is exactly compensated by the injection of atoms in 
the annular region, so that the number of atoms in the system 
remains constantly equal to the initial atom number.

It is interesting to relate $V_0$ (the amplitude of the imaginary potential 
in Eq.~(\ref{eq:trap}) generating the drain flow) to the the quantity
which is controlled in the experiment: the flow rate from the injection annulus
into the drain hole, $Q_{inj}$.
To do this, we integrate Eq.~(\ref{eq:cont}) over the volume ${\cal V}_s$
of the drain hole, and define

\begin{equation}
Q_{sink}=\oint_{{\Sigma}_s} n \boldv \cdot d\boldS,
\end{equation}

\noindent
the flow rate out of ${\cal V}_s$, where ${{\Sigma}_s}$ is the surface 
which encloses ${\cal V}_s$. Because of the box-trap boundary conditions, 
the only flow in and out of ${\cal V}_s$ is across 
the top surface $\overline{\Sigma}_s$ of the truncated cone, {\it i.e.}
$\displaystyle Q_{sink}=\int_{\overline{\Sigma}_s} n \boldv \cdot d\boldS$.
We find 

\begin{equation}
\frac{dN_{sink}}{dt}+Q_{sink}=-2 V_0 N_{sink},
\end{equation}

\noindent
In the steady state $Q_{inj}=-Q_{sink}$, hence

\begin{equation}
Q_{inj}=2 N^0_{sink} V_0.
\label{eq:Qinj}
\end{equation}

\noindent
The steady number of atoms in the drain hole $N^0_{sink}$ is not
constant but depends on the potential $V_0$, 
implying that the relation between $Q_{inj}$ and
$V_0$, Eq.~(\ref{eq:Qinj}), is not linear. The dependence of $N^0_{sink}$ on
$V_0$ can be, at least qualitatively, understood employing the Bernoulli 
equation which can be derived \cite{Primer} from
Eq.~(\ref{eq:GPE}) neglecting the quantum pressure effects:

\begin{equation}
\frac{1}{2} v_s^2+ n_s=
\frac{1}{2} v_c^2+ n_c \, ,
\label{eq:Berno}
\end{equation}

\noindent
where $v_s$, $p_s$ and $v_c$, $p_c$ are velocities and pressures
respectively in the centre of the drain hole just below the cylinder 
(where the velocity is $\boldv=-v_s \zhat$, $\zhat$ being the unit vector
in the $z$-direction and $v_s > 0$) and in the cylinder far away
from the drain hole (where $|\boldv| = v_c \approx 0$). 
Employing the Bernoulli equation Eq.~(\ref{eq:Berno}), we obtain
\begin{equation}
\displaystyle
v_s=\sqrt{2 \left ( n_c - n_s \right )},
\label{eq:Berno_v}
\end{equation}
\noindent
where the density in the drain hole is approximately
$n_s=N^0_{sink}/{\cal V}_s$. %, $N^0_{sink}$
%being the steady state number of atoms in the drain hole. 
In first approximation, the flux across the top surface $\overline{\Sigma}_s$ of the drain hole is 
$Q_{inj}=n_s v_s A_s$, where $A_s$ is the area of $\overline{\Sigma}_s$.
Using this last relation, the continuity equation Eq.~(\ref{eq:Qinj}) and the Bernoulli relation 
Eq.~(\ref{eq:Berno_v}), we obtain the following expression for $N^0_{sink}$:

\begin{equation}
N^0_{sink}=n_c {\cal V}_s -\frac{2 V_0^2 {\cal V}_s^3}{A_s^2} \, ,
\end{equation}

\noindent
leading to 

\begin{equation}
Q_{inj}=2 V_0 {\cal V}_s (n_c - \frac{2 V_0^2 {\cal V}_s^2}{S_s^2})\, .
\label{eq:q_inj_nonlin}
\end{equation}

\begin{figure}[!ht]
\includegraphics[angle=0,width=0.8\columnwidth]{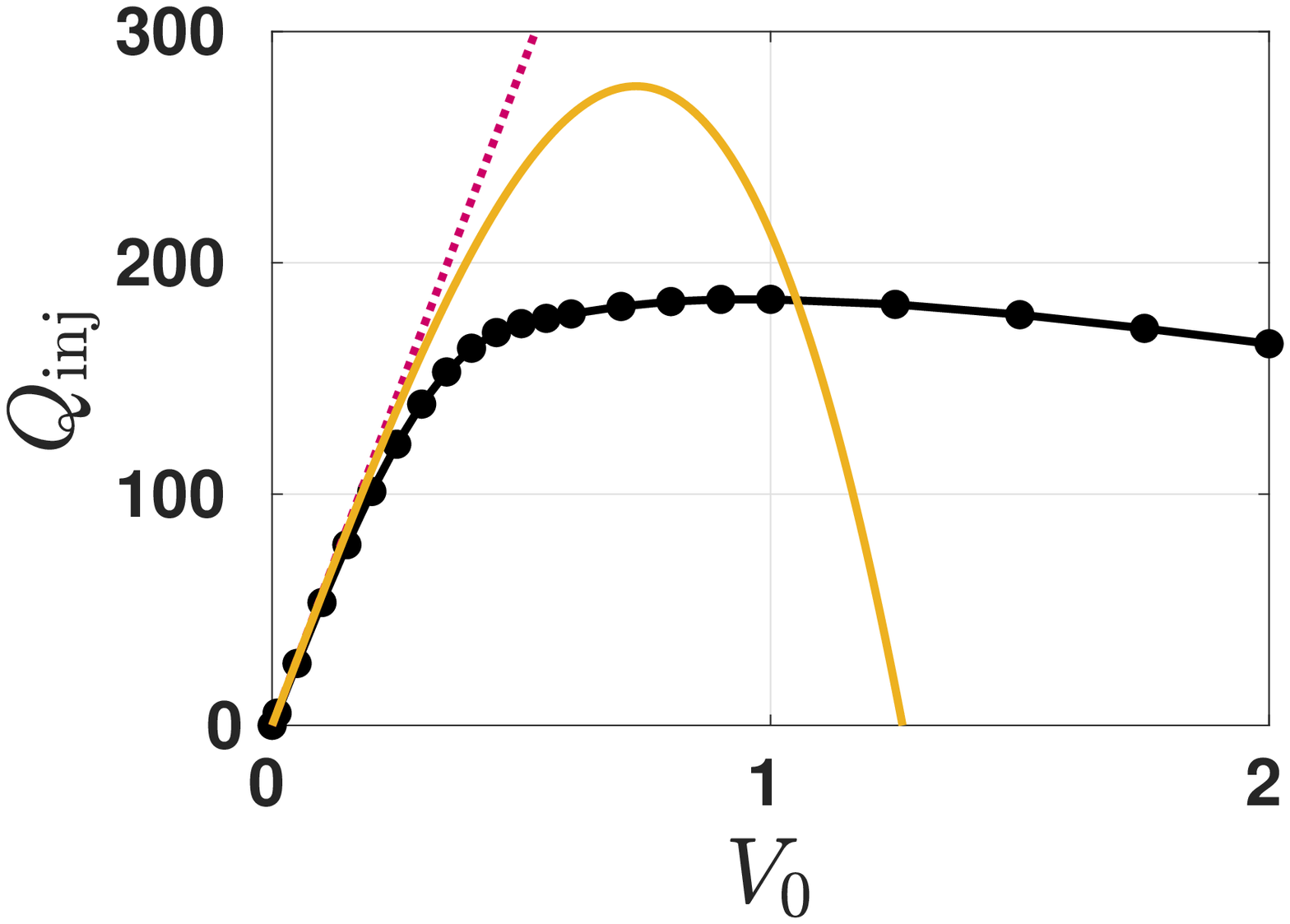}\\
\includegraphics[angle=0,width=0.8\columnwidth]{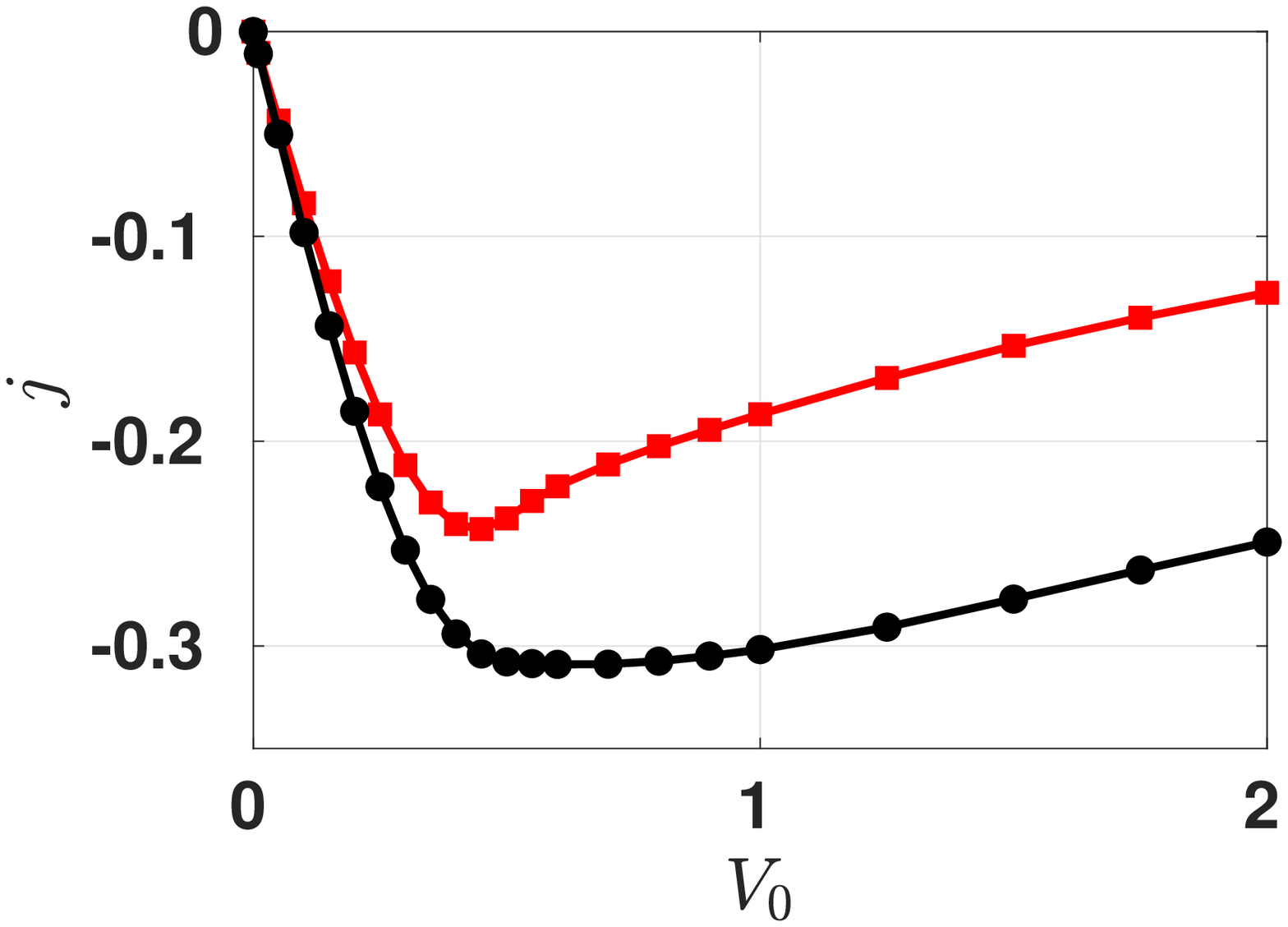}
\caption{
Top (a): steady-state flow rate out of the injection annulus 
into the cylinder, $Q_{inj}$, as a function of  
the amplitude of the complex trapping potential $V_0$ 
which generates the drain flow (solid black line). 
The solid yellow line corresponds to Eq.~(\ref{eq:q_inj_nonlin}), 
while the dashed red line shows its linear
approximation at small values of $V_0$.
Bottom (b): steady-state averaged radial and axial current, 
$\overline{j_r}$ (solid red line) and $\overline{j_z}$ (solid black line) 
respectively, as a function of $V_0$.
The parameters are: $N_x=N_y=140$, $N_z=120$, $x_{min}=y_{min}=-35$, 
$x_{max}=y_{max}=35$, $z_{min}=-30$, $z_{max}=30$,
$z_1=-28$, $z_2=29$, $z_3=-28$, $z_4=-24$, $r_{ad}=31$, $r_{in}=34$,
$r_{sink}=3$, $r'_{sink}=12$.
}
\label{fig:2}
\end{figure}

\noindent
This non-linear relation between $Q_{inj}$ and $V_0$ can be observed 
in Fig.~\ref{fig:2}(a)
where we plot Eq.~(\ref{eq:q_inj_nonlin}) and its linear approximation at small $V_0$.
It is clear that our assumptions in deriving Eq.~(\ref{eq:q_inj_nonlin}) are valid when 
$V_0$ is not too large compared to the chemical potential. 

\noindent
This nonlinear effect is also visible in Fig.~\ref{fig:2}(b) 
where we plot the dependence on the potential $V_0$ of 
the averaged radial and axial components of the current, 
$\overline{j_r}$ and $\overline{j_z}$
respectively, flowing into the drain hole,
averaged over a horizontal disk of radius $r'_{sink}$ at $z=(z_3+z_4)/2$.

\begin{figure}[htbp]
\centering
\includegraphics[angle=0,width=0.95\columnwidth]{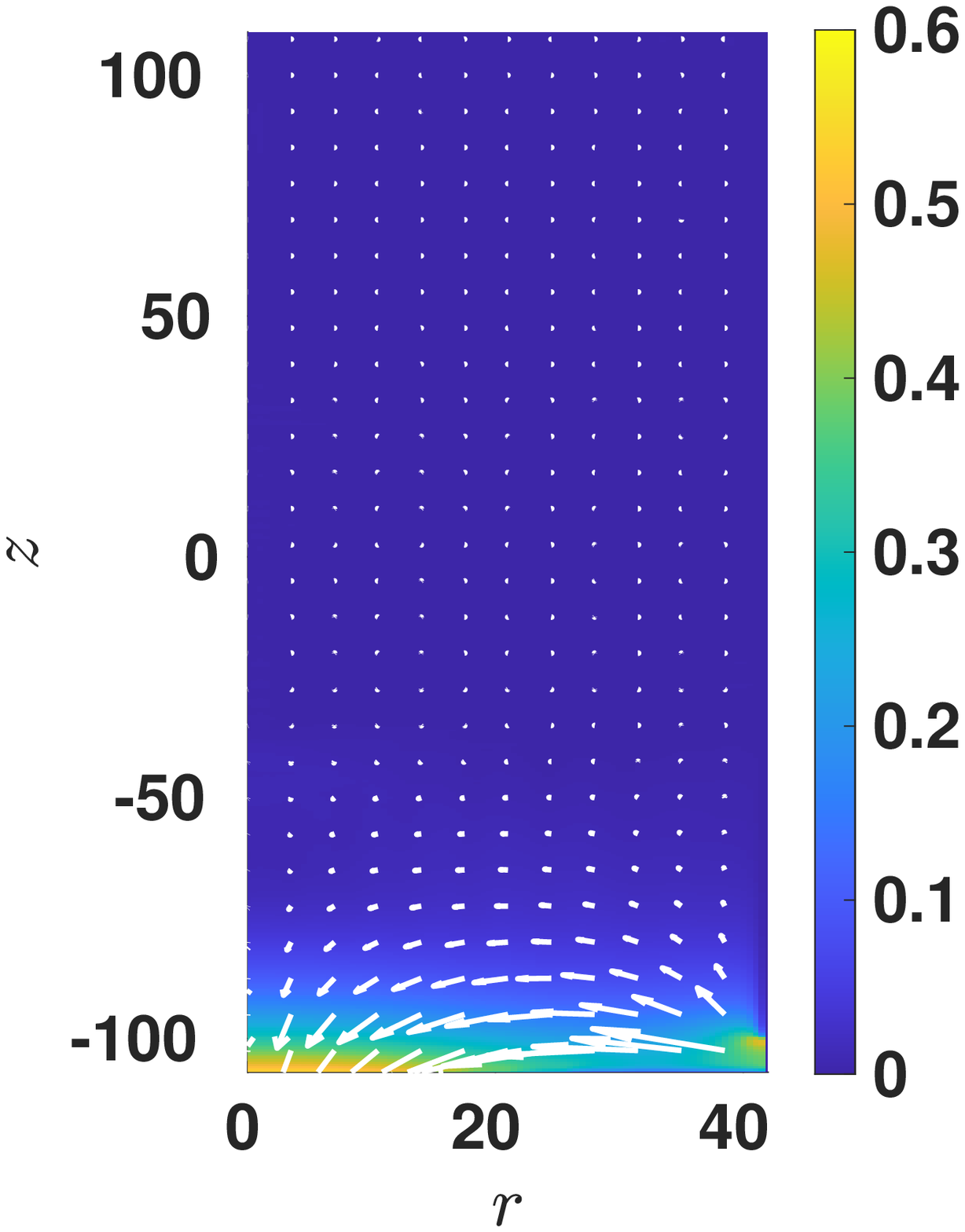}\\
\includegraphics[angle=0,width=0.85\columnwidth]{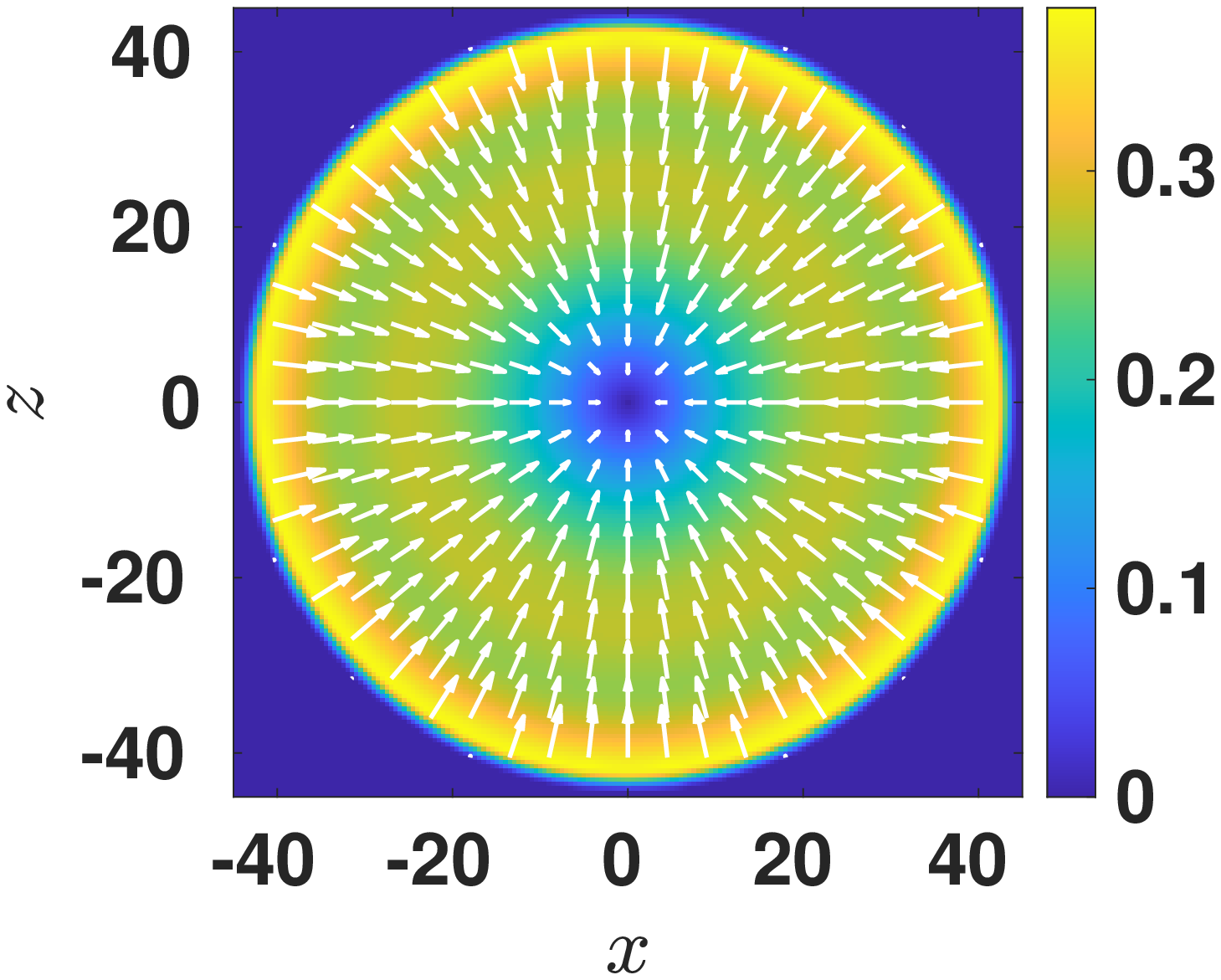}
\caption{
Top (a): Steady drain flow pattern (in the absence of vortex lines)
plotted on the $xz$ plane for $y=0$. The arrows indicate the 
direction of the current and the colours represent the current's magnitude.
For clarity, only the flow in the cylinder is plotted,
ignoring the drain hole and the injection annulus.
Bottom (b): Steady drain flow pattern similarly
plotted on the $xy$ plane at $z=(z_3+z_4)/2$.
The parameters are: $N_x=N_y=180$, $N_z=440$, $x_{min}=y_{min}=-45$,
$x_{max}=y_{max}=45$, 
$z_{min}-110$, 
$z_{max}=110$, $z_1=-107$,
$z_2=109$, $z_3=z_1$, $z_4=z_1+7$,
$r_{ad}=31$, $r_{in}=42$, $r_{sink}=5$, $r'_{sink}=30$ and $V_{sink}^0=0.1$.
}
\label{fig:3}
\end{figure}

\begin{figure}[htbp]
\centering
\includegraphics[angle=0,width=0.9\columnwidth]{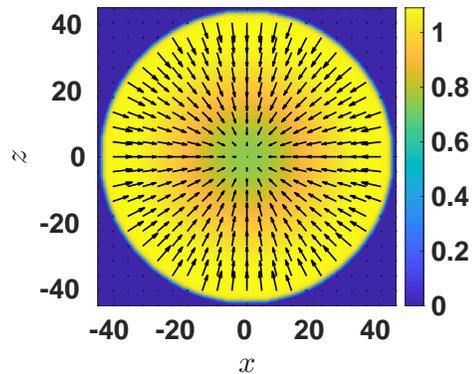}
\caption{
Steady drain flow pattern 
plotted on the $xy$ plane at $z=(z_3+z_4)/2$. 
The arrows indicate the 
direction of the current and the colours 
correspond to pressure values. Parameters as in Fig.~\ref{fig:3}.
}
\label{fig:quiver_pressure}
\end{figure}

Fig.~\ref{fig:3} shows the steady drain flow pattern in a larger geometry
plotted on the $xz$ plane (a) and on the $xy$ plane (b). 
It is apparent that the radial and axial flows
into the drain hole are confined to the bottom of the cylinder, the radial
component being stronger than the axial one. This flow pattern is very different
from the flow pattern of a viscous drain hole described in
Ref.~\cite{Andersen}, which suffers viscous friction near all boundaries;
our drain flow is instead inviscid and irrotational.
The plot of the pressure distribution (see Fig.~\ref{fig:quiver_pressure})
confirms that the flow is driven by pressure gradients, as expected.

%Moreover, at
%small velocity, the density variations in our flow are small, so
%the solution which we have found is qualitatively similar to the 
%solution of Laplace equation for an irrotational incompressible fluid.

\subsection{\label{sec:vortexflow}Drain vortex flow}
%\label{sec:vortexflow}

In the next numerical experiment, we compute the time
evolution of a lattice of vortex lines in the presence of a
drain flow.  
%The parameters which we use are
%$z_1=-107$, $z_2=109$, $z_3=z_1=-107$, $z_4=z_1+7=-100$,
%$r_{ad}=42$, $r_{in}=44.5$, $r_{sink}=5$, $r'_{sink}=30$, and
%$V_{sink}^0=0.1$ 
%\red{[Curiosity: how much is $z_{min}$?][$z_{min}=-110$]}.%
We add $N_v=10$ parallel vortex lines aligned along the $z$-direction
which form a small lattice (bundle) of initial radius $r_b=25$.
This vortex imprinting is done in a standard way
by multiplying the ground state wavefunction with the approximate 
wavefunction of a vortex in a homogeneous condensate and let 
the system then relax in imaginary time.

\begin{figure*}[!ht]
     \begin{minipage}{0.15\textwidth}
      \centering
       \includegraphics[width=1\textwidth]{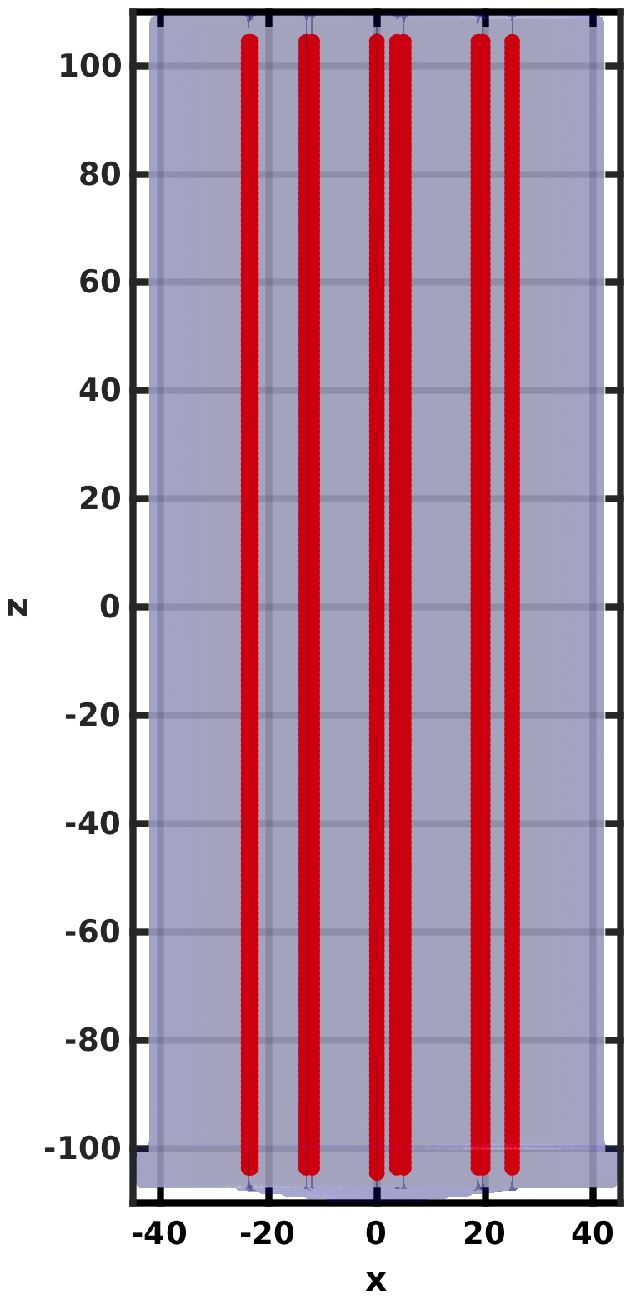}
     \end{minipage}
    \hspace{0.008\textwidth}
     \begin{minipage}{0.15\textwidth}
      \centering
       \includegraphics[width=1\textwidth]{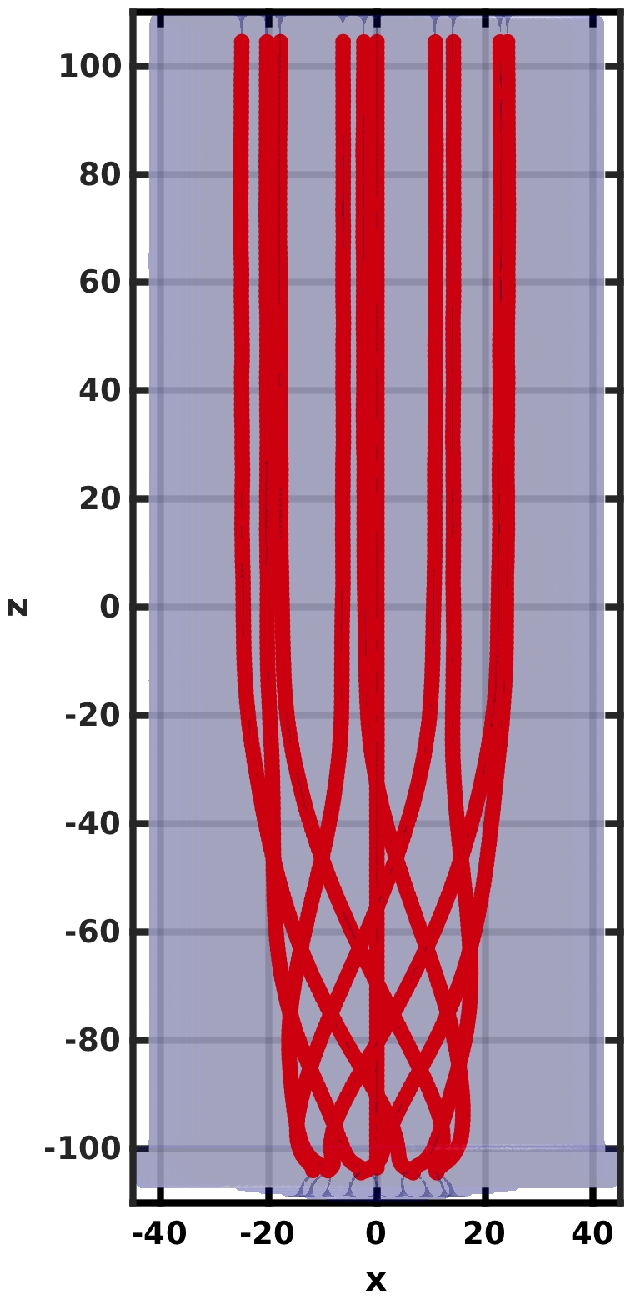}
      \end{minipage}
    \hspace{0.008\textwidth}
     \begin{minipage}{0.15\textwidth}
      \centering
       \includegraphics[width=1\textwidth]{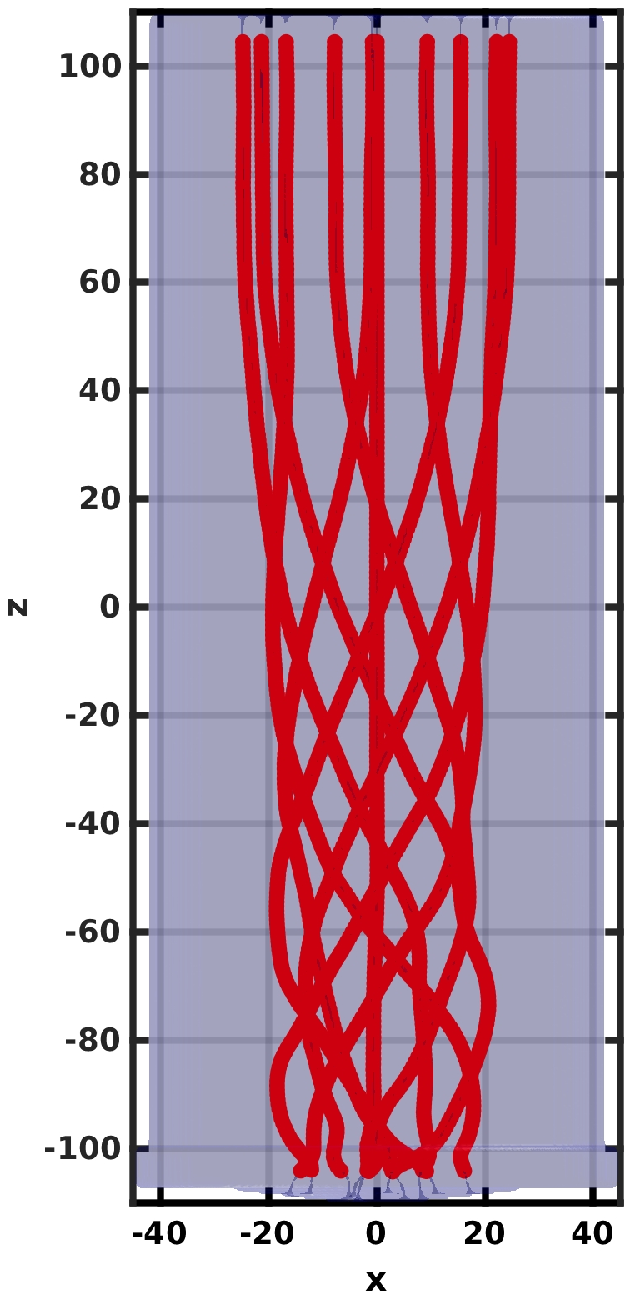}
     \end{minipage}
    \hspace{0.008\textwidth}
      \begin{minipage}{0.15\textwidth}
      \centering
       \includegraphics[width=1\textwidth]{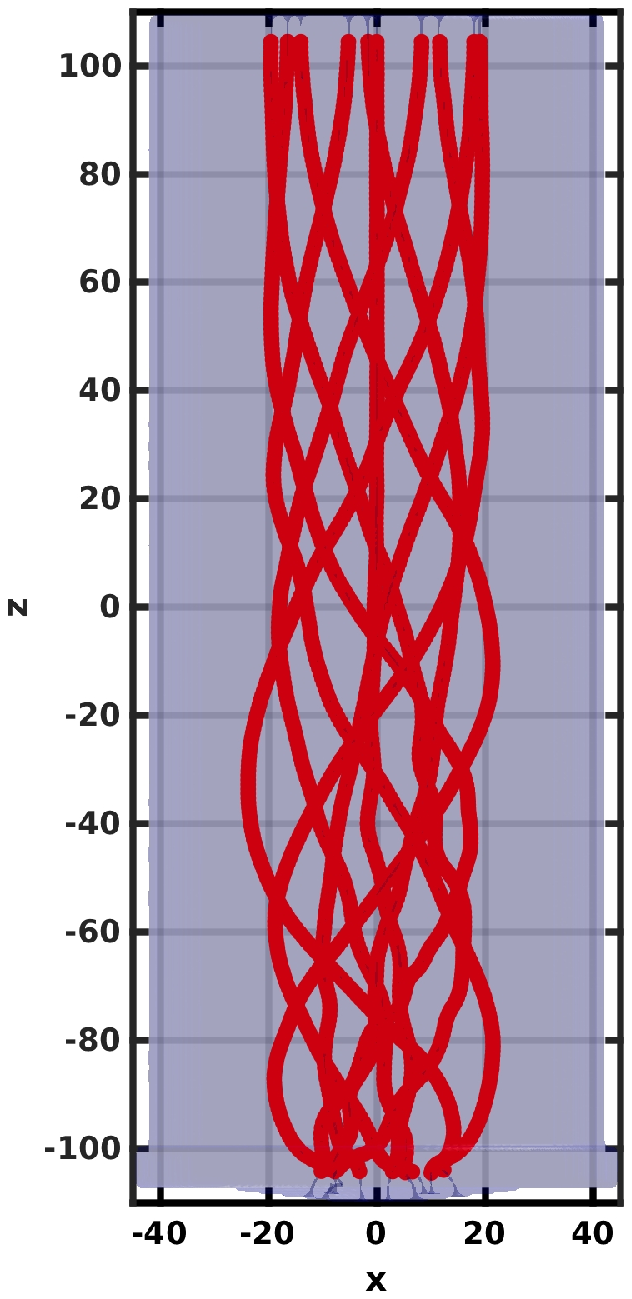}
      \end{minipage}
    \hspace{0.008\textwidth}
      \begin{minipage}{0.15\textwidth}
      \centering
       \includegraphics[width=1\textwidth]{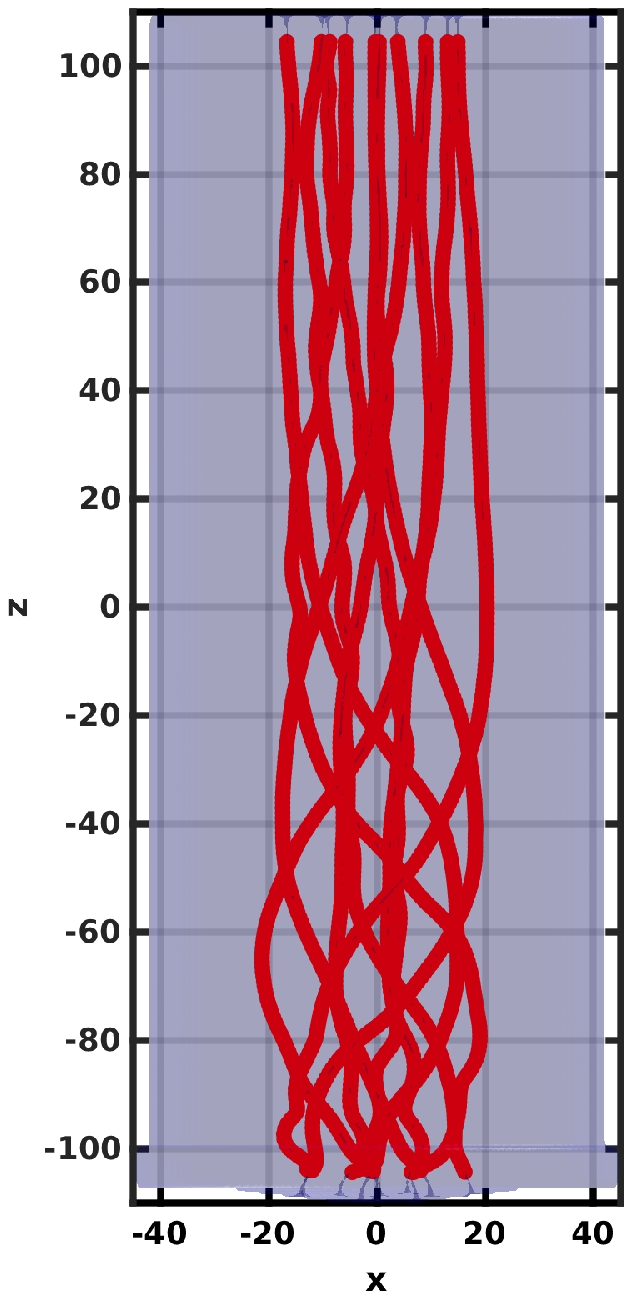}
      \end{minipage}
     \hspace{0.008\textwidth}
      \begin{minipage}{0.15\textwidth}
      \centering
       \includegraphics[width=1\textwidth]{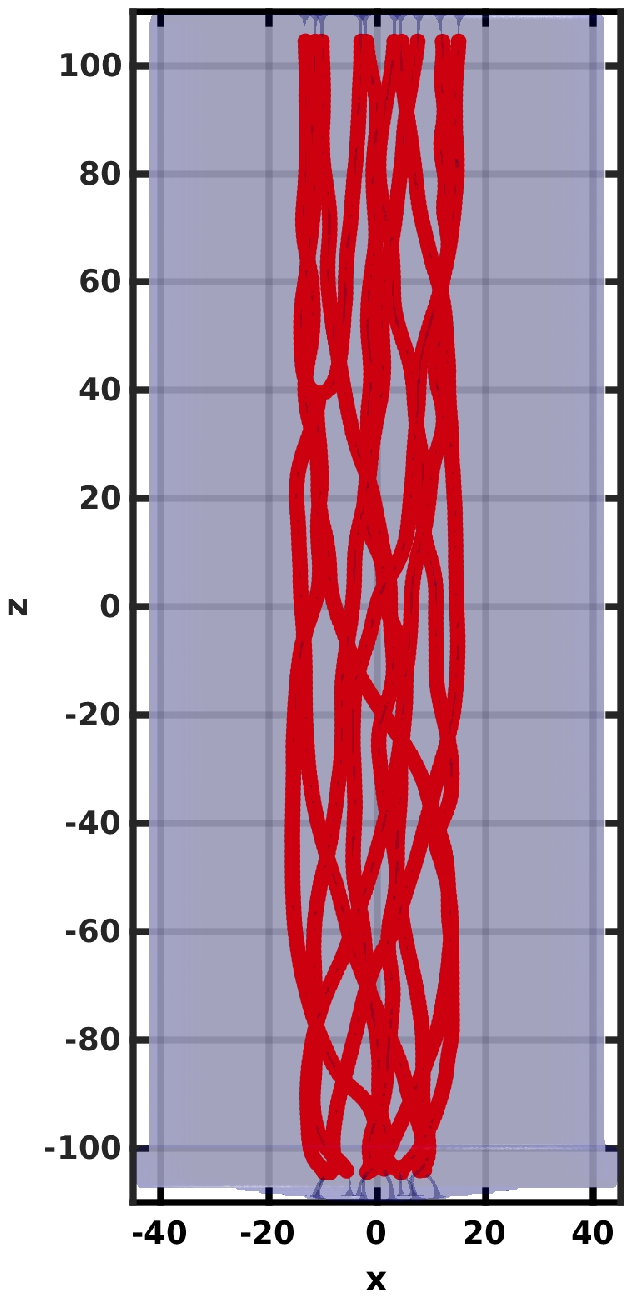}
      \end{minipage}
     %\hspace{0.007\textwidth}
      %\begin{minipage}{0.2\textwidth}
      %\centering
      % \includegraphics[width=0.8\textwidth]{./BEC_r0_022_022_z0_5_-5_yx/tracks/delta_mean.ps}
      %\end{minipage}
      \\%[2mm]
%%%%%%%%%%%%%%%%%%%%%%%%%%%%%%%%%%%%%%%
     \hspace{0.00675\textwidth}
     \begin{minipage}{0.15\textwidth}
      \centering
       \includegraphics[width=0.9\textwidth]{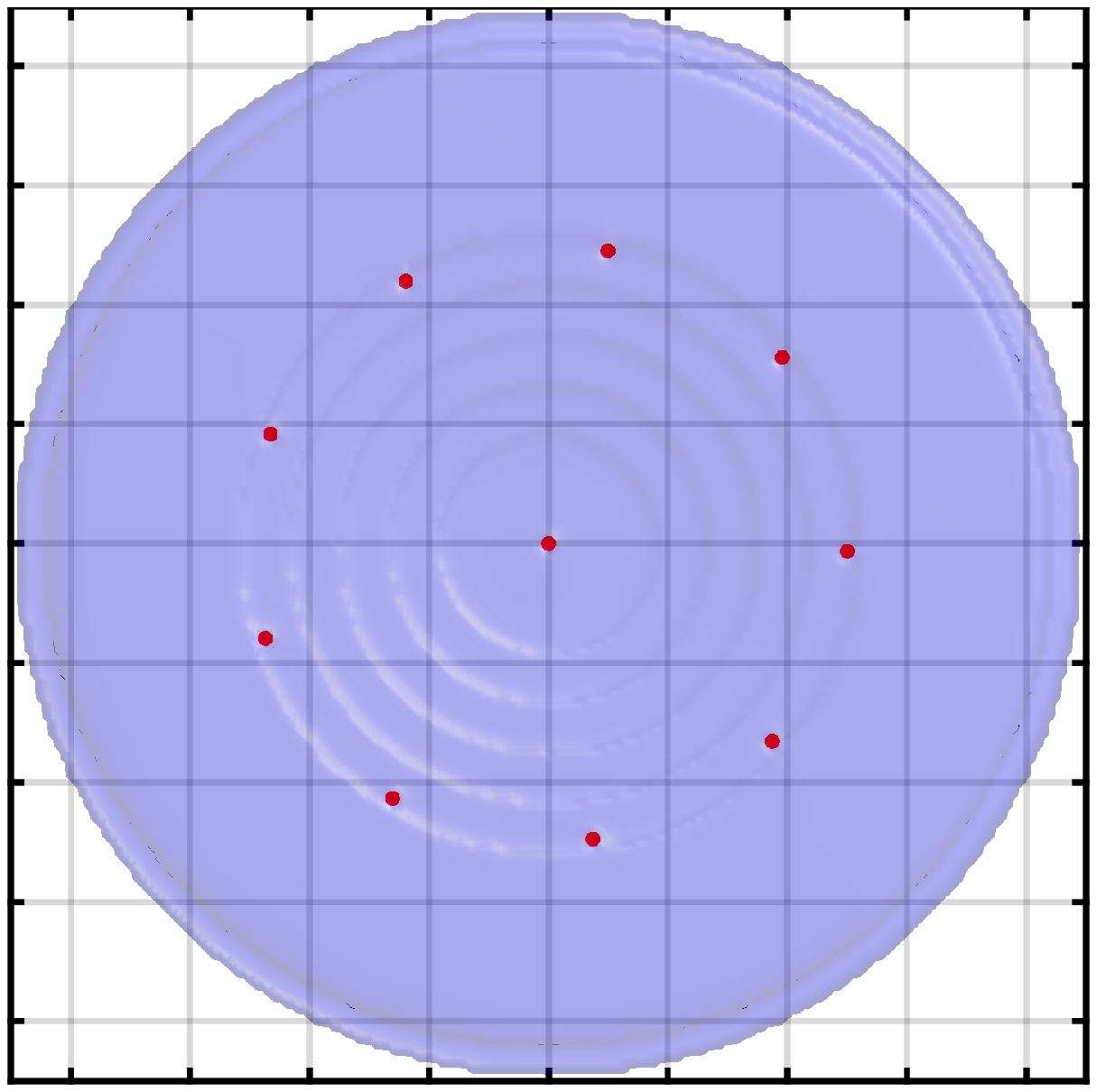}
     \end{minipage}
    \hspace{0.00695\textwidth}
     \begin{minipage}{0.15\textwidth}
      \centering
       \includegraphics[width=0.9\textwidth]{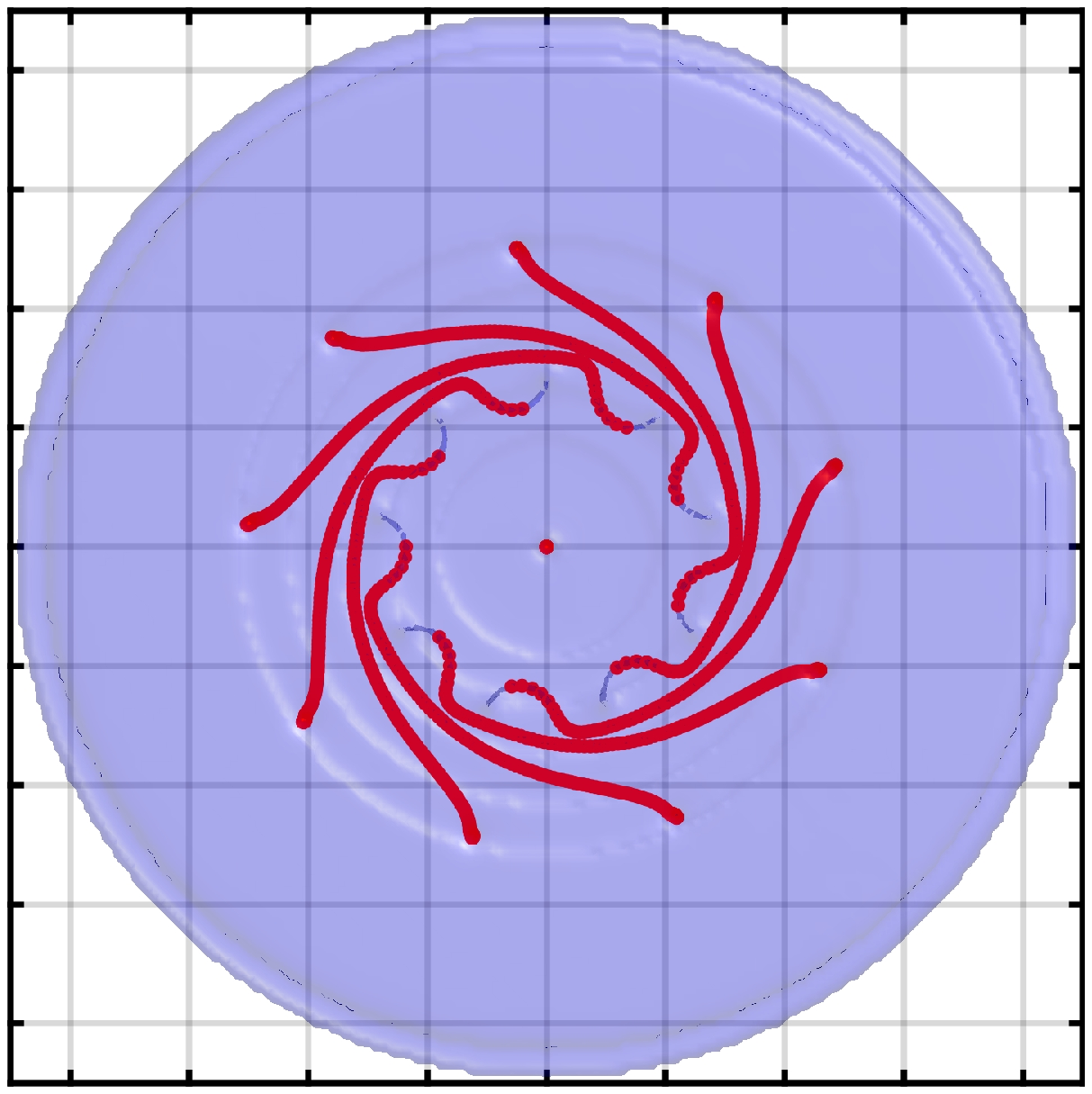}
      \end{minipage}
    \hspace{0.00695\textwidth}
     \begin{minipage}{0.15\textwidth}
      \centering
       \includegraphics[width=0.9\textwidth]{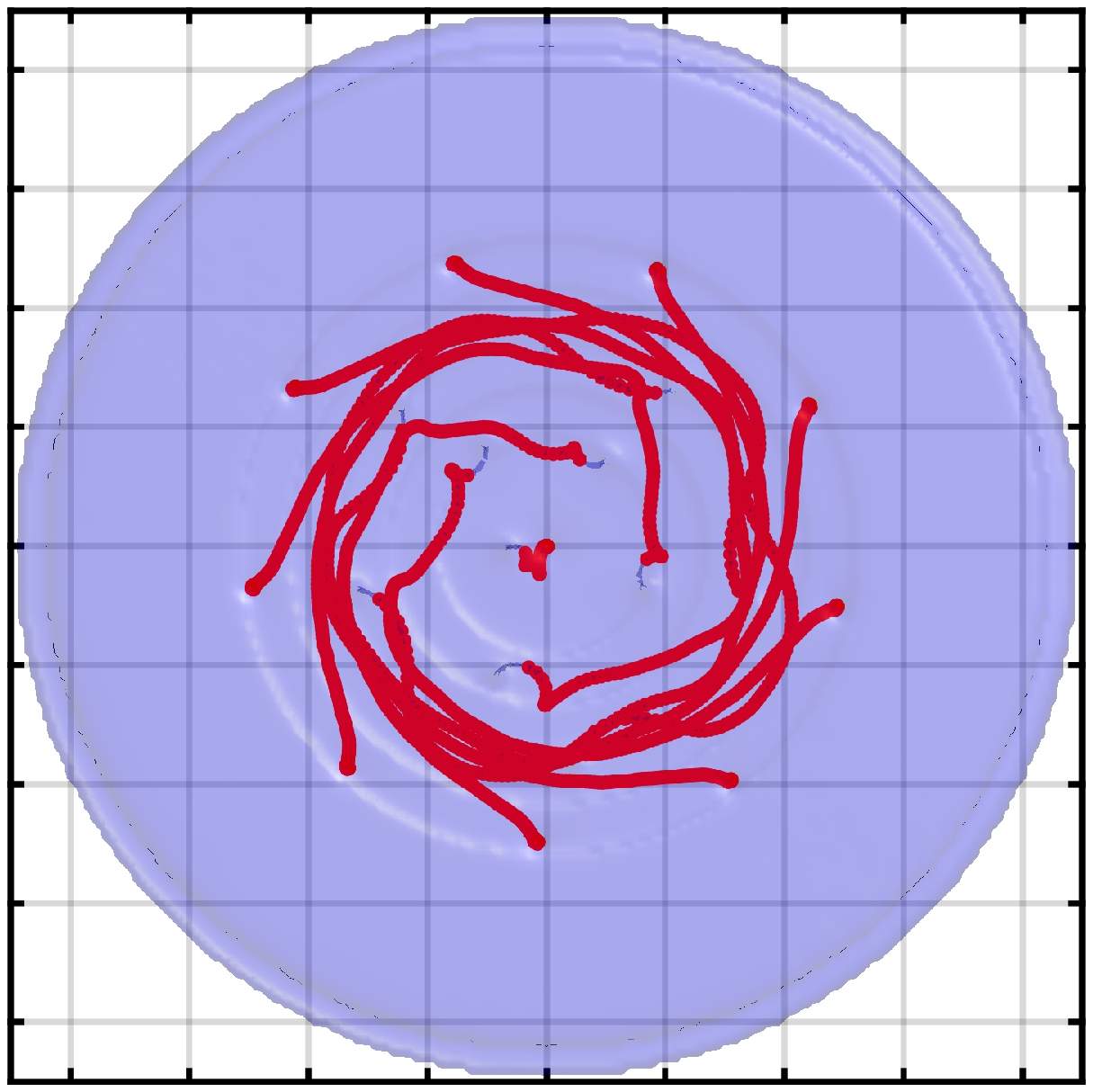}
     \end{minipage}
    \hspace{0.00685\textwidth}
      \begin{minipage}{0.15\textwidth}
      \centering
       \includegraphics[width=0.9\textwidth]{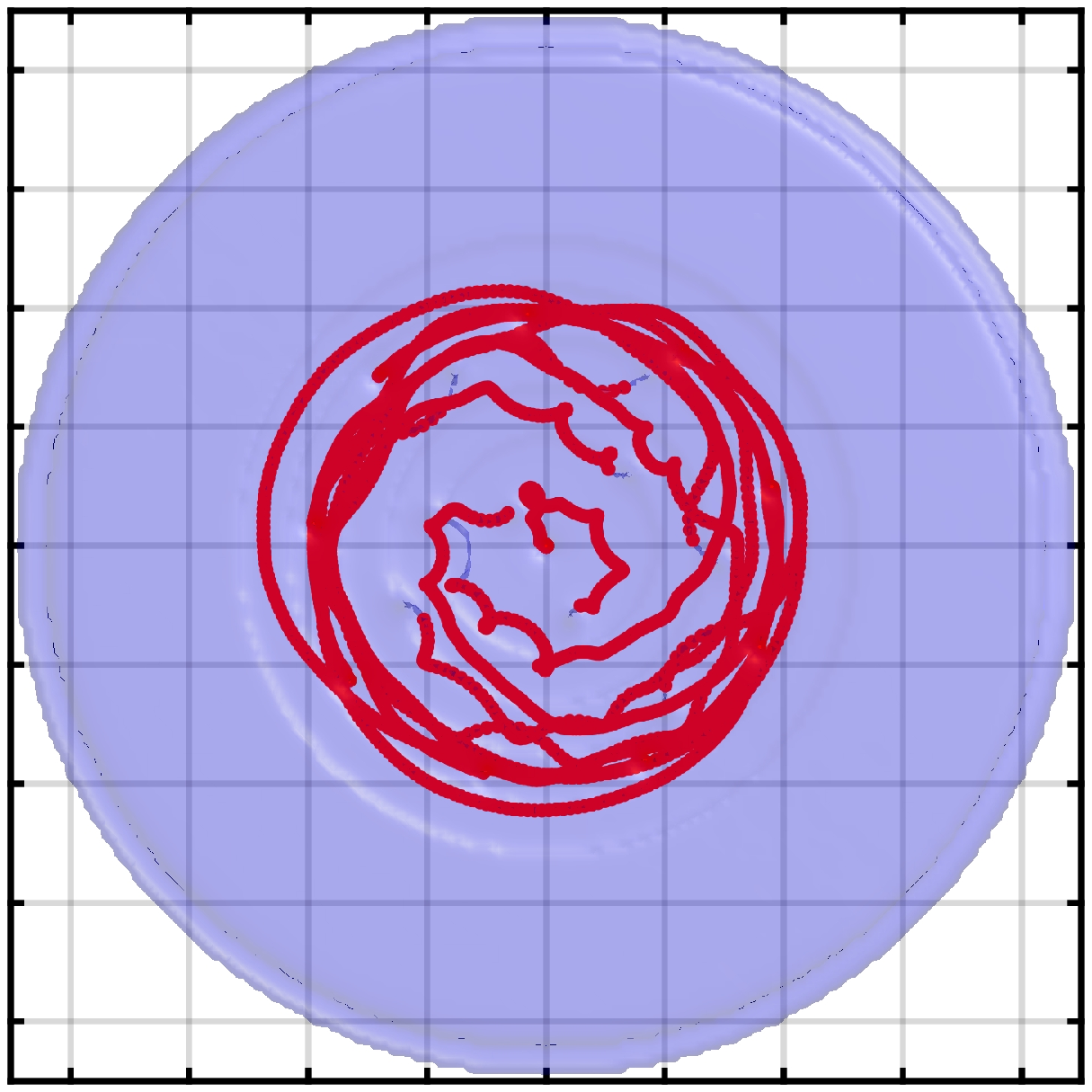}
      \end{minipage}
    \hspace{0.00685\textwidth}
      \begin{minipage}{0.15\textwidth}
      \centering
       \includegraphics[width=0.9\textwidth]{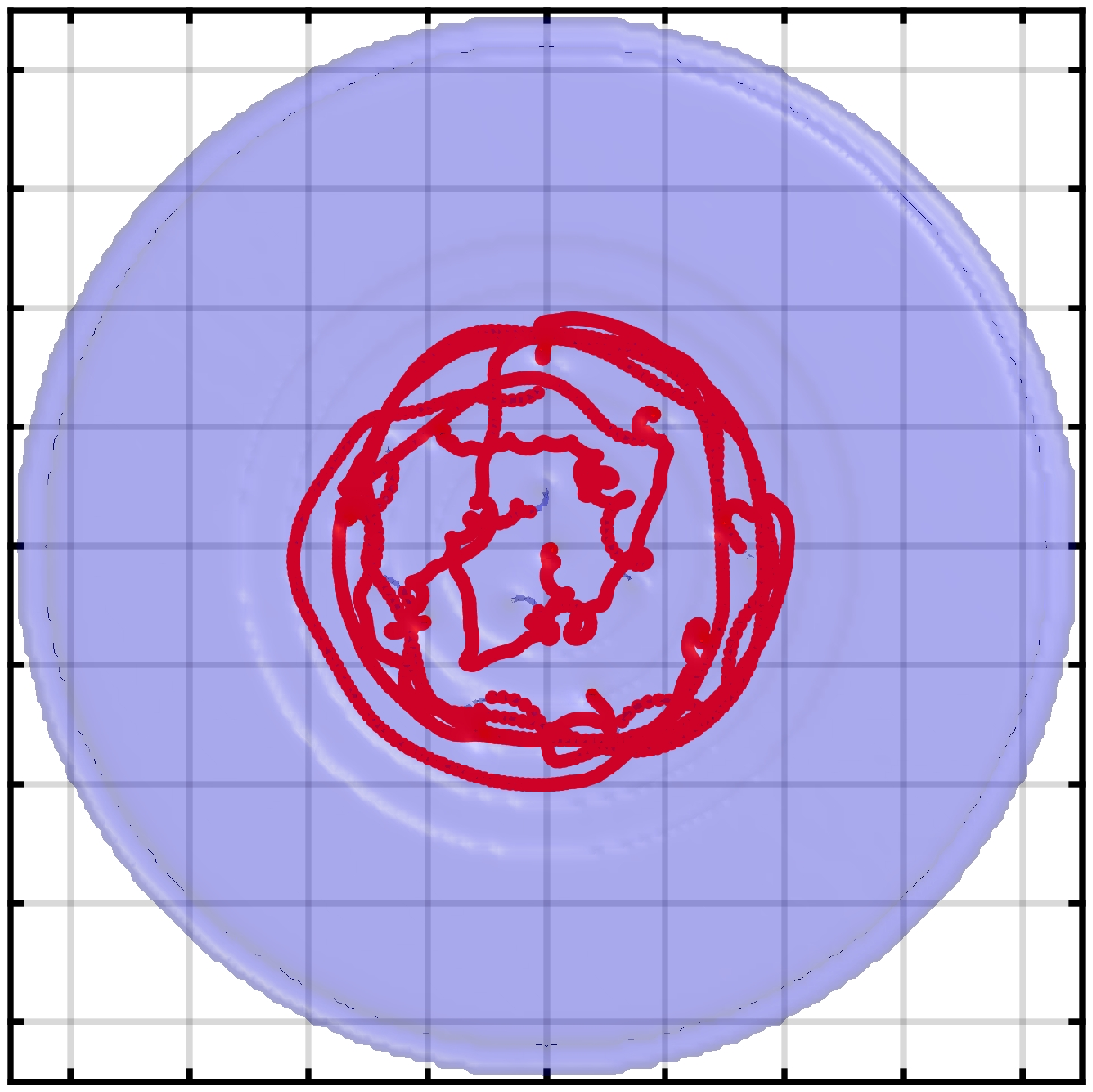}
      \end{minipage}
     \hspace{0.00695\textwidth}
      \begin{minipage}{0.15\textwidth}
      \centering
       \includegraphics[width=0.9\textwidth]{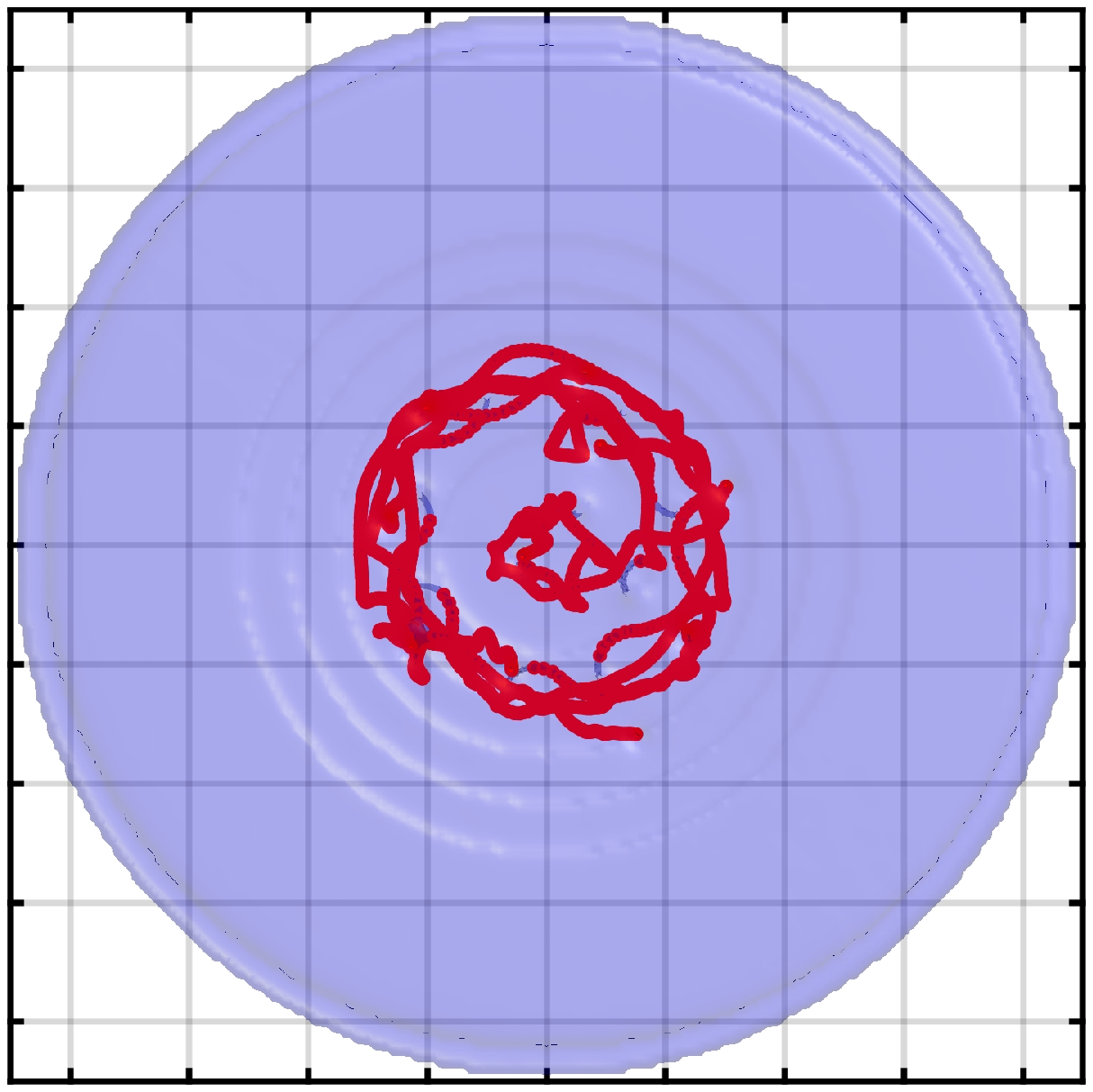}
      \end{minipage}
\caption{Snapshots of the vortex configurations at times
$t=0,\, 320,\, 640,\, 
960,\, 1280,\, 1600$. The parameters are as in Fig.~\ref{fig:3}.
}      
\label{fig:6}
\end{figure*}

\begin{figure*}[!ht]
\centering
\includegraphics[angle=0,width=0.8\columnwidth]{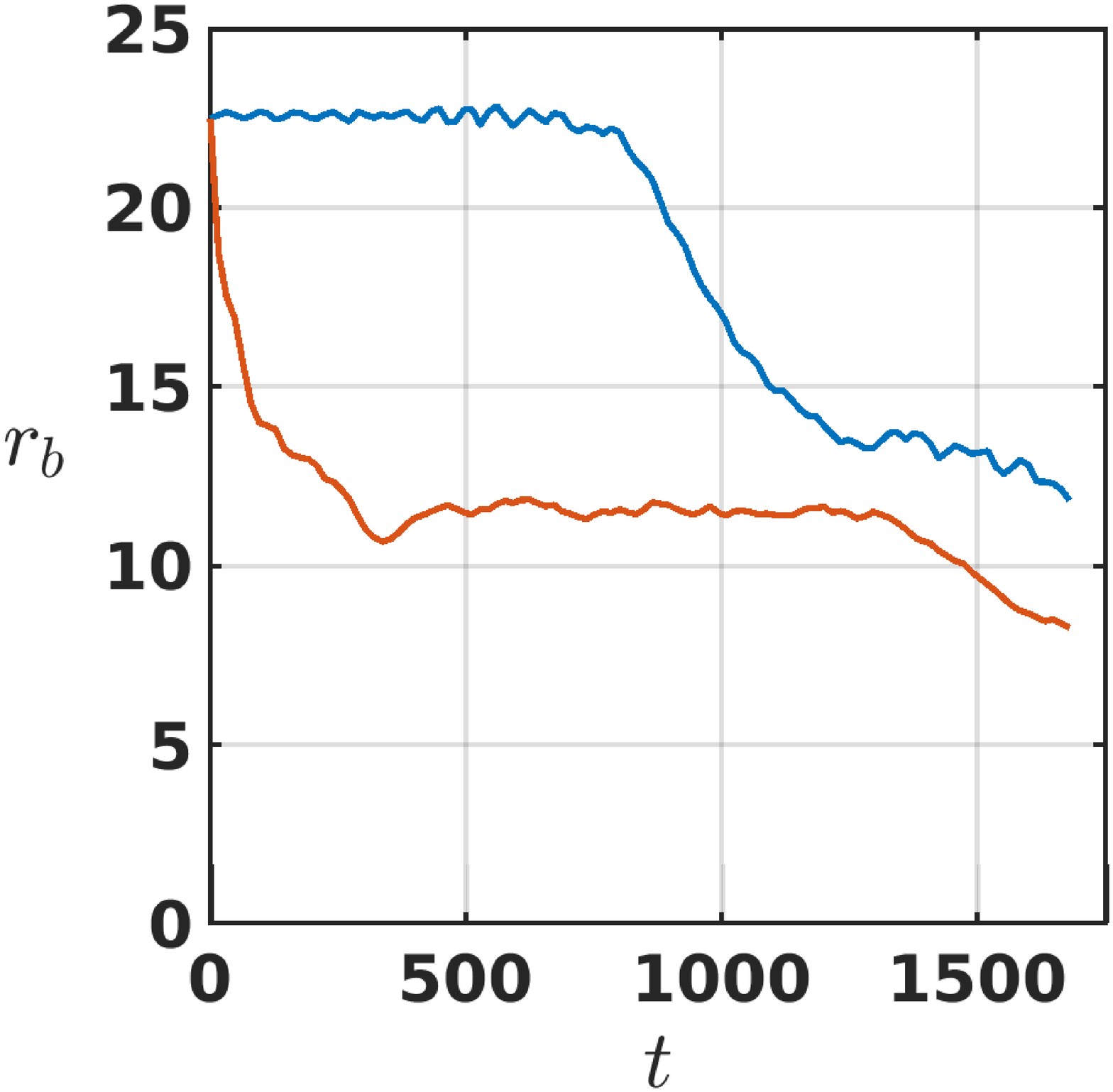}
\hspace{0.1\columnwidth}
\includegraphics[angle=0,width=0.8\columnwidth]{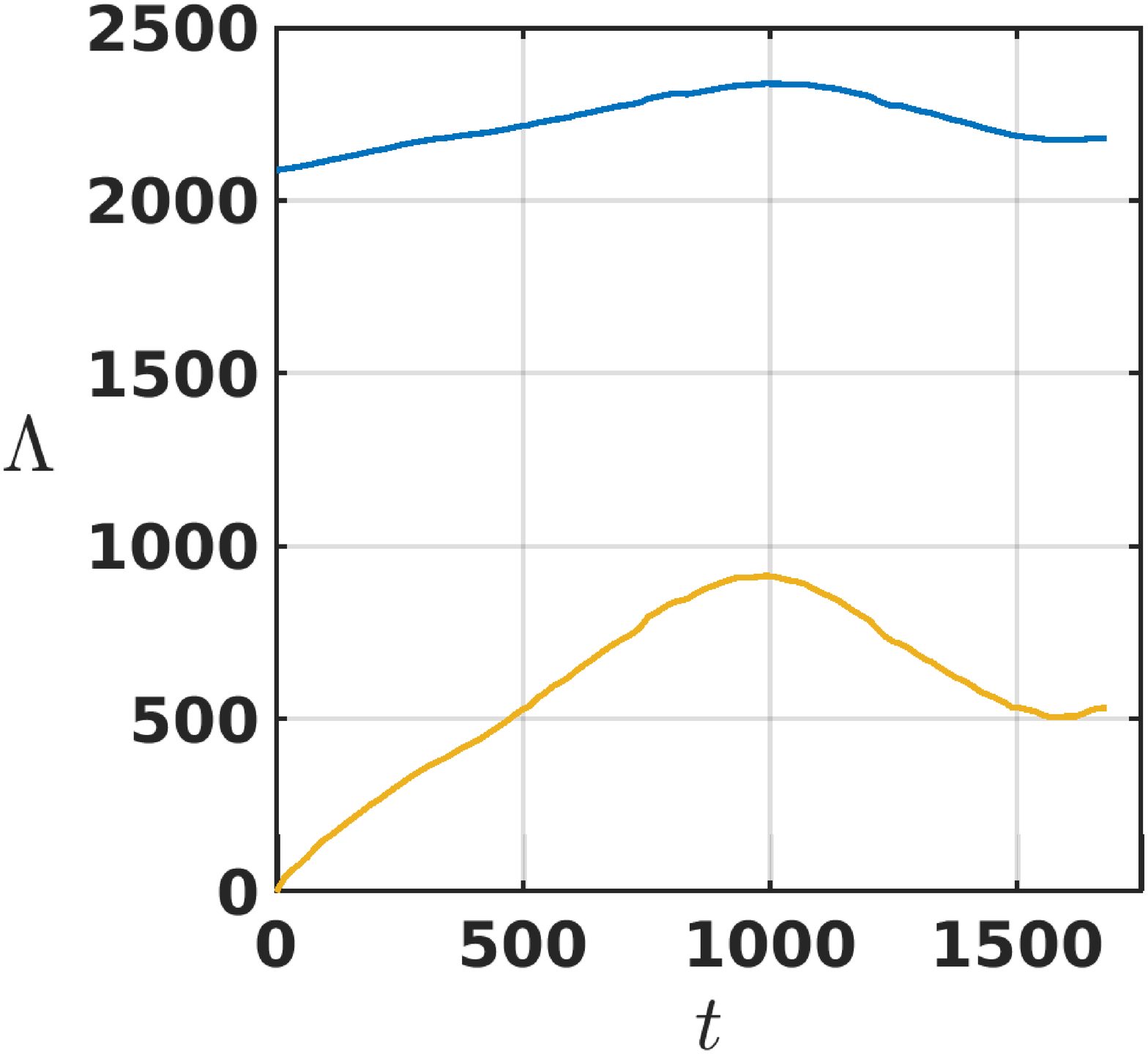}
\caption{
Left (a): temporal evolution of the top (blue) and bottom (red)
average radius of the vortex bundle $r_b$;
Right (b): temporal evolution of the total length 
$\Lambda$ of the vortex bundle
(blue) and its orthogonal projection $\Lambda_\perp$ on a plane perpendicular
to the cylinder's axis (yellow). 
The parameters are as in Fig.~\ref{fig:3}.
}
\label{fig:lenght_rb}
\end{figure*}

By integrating in time the GPE, we follow the evolution of the vortex
bundle, which is shown in Fig.~\ref{fig:6}. 
We find that, since vortex lines are advected by the superflow
(Helmholtz's theorem) and since near the bottom of the cylinder
the inward radial flow is much stronger than near the top, 
the lower part of the bundle is sucked radially inwards, towards the 
center of the drain hole, while the top
part is basically unaffected (this effect is also apparent in
Fig.~\ref{fig:lenght_rb} (a) which displays the bundle's radius at two
different heights). 
The quantisation of the superfluid's angular
momentum \cite{Feynman,Primer} implies that the angular velocity $\Omega (z,t)$
of the bundle at height $z$ and time $t$ is given by the relation
$\Omega(z,t)=N_v\kappa/[2\pi r_b^2(z,t)]$. This implies that 
the bottom parts of the vortex
lines which are near the drain hole rotate at larger angular 
velocity $\Omega$ (as $r_b$ is smaller), moving hence ''ahead'' 
with respect to
the top parts of the vortex lines close to the top {of the
cylinder}: the vortex bundle becomes more and more twisted,
as shown in Fig.~\ref{fig:6} (b). 
As this twist develops, the projection of the total vortex 
length on a plane orthogonal to the cylinder axis, $\Lambda_\perp$, 
increases, leading to an increase of the total length, $\Lambda$,
as shown in Fig.~\ref{fig:lenght_rb} (b). 
It takes a finite time for this
initial twist to propagate upwards (see Fig.~(\ref{fig:6})) carried by 
Kelvin waves on vortices. Once the Kelvin waves reach the top of the
cylinder, the bundle's radius, $r_b$, starts decreasing 
near the top as well, decreasing 
$\Lambda_\perp$. However, given the enduring asymmetry of the radial flow 
(stronger close to the drain hole, negligible near the top of the cylinder),
Kelvin waves persist on the vortices, leading $\Lambda_\perp$ to 
settle to a finite, non-zero value. The dynamics of this rotating, 
twisted bundle is characterised by vortex reconnections
which occasionally scramble the vortex lines, leading to a 
moderately disordered vortex configuration which is still strongly 
polarised in the $z$ direction.
%%%%%%%%%%%%%%%%%%%%%%%%%%%%%%%%%%%

%%%%%%%%%%%%%%%%%%%%%%%%%%%%%%%%

\begin{figure}[!ht]
\centering
\includegraphics[angle=0,width=0.85\columnwidth]{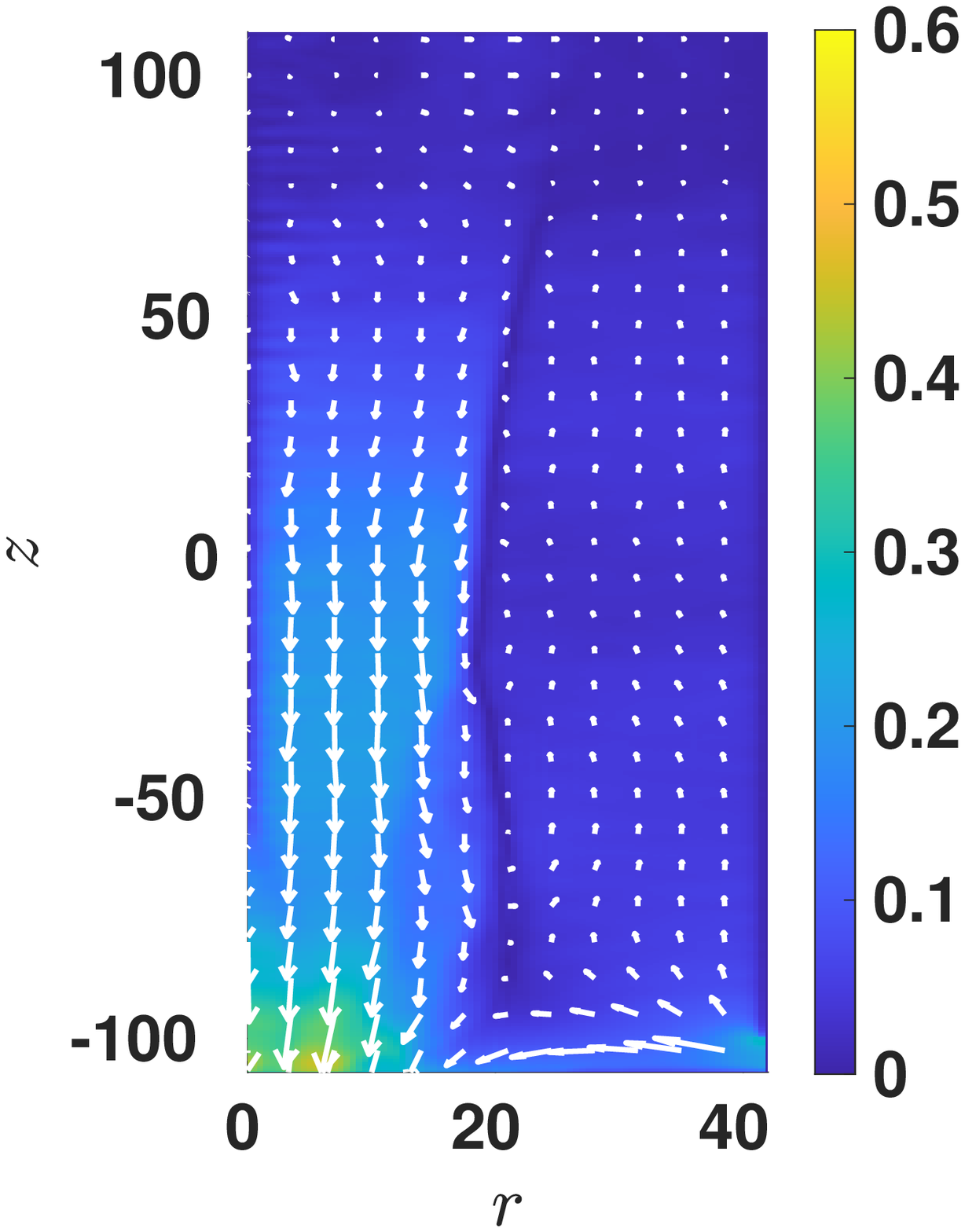}\\
\includegraphics[angle=0,width=0.95\columnwidth]{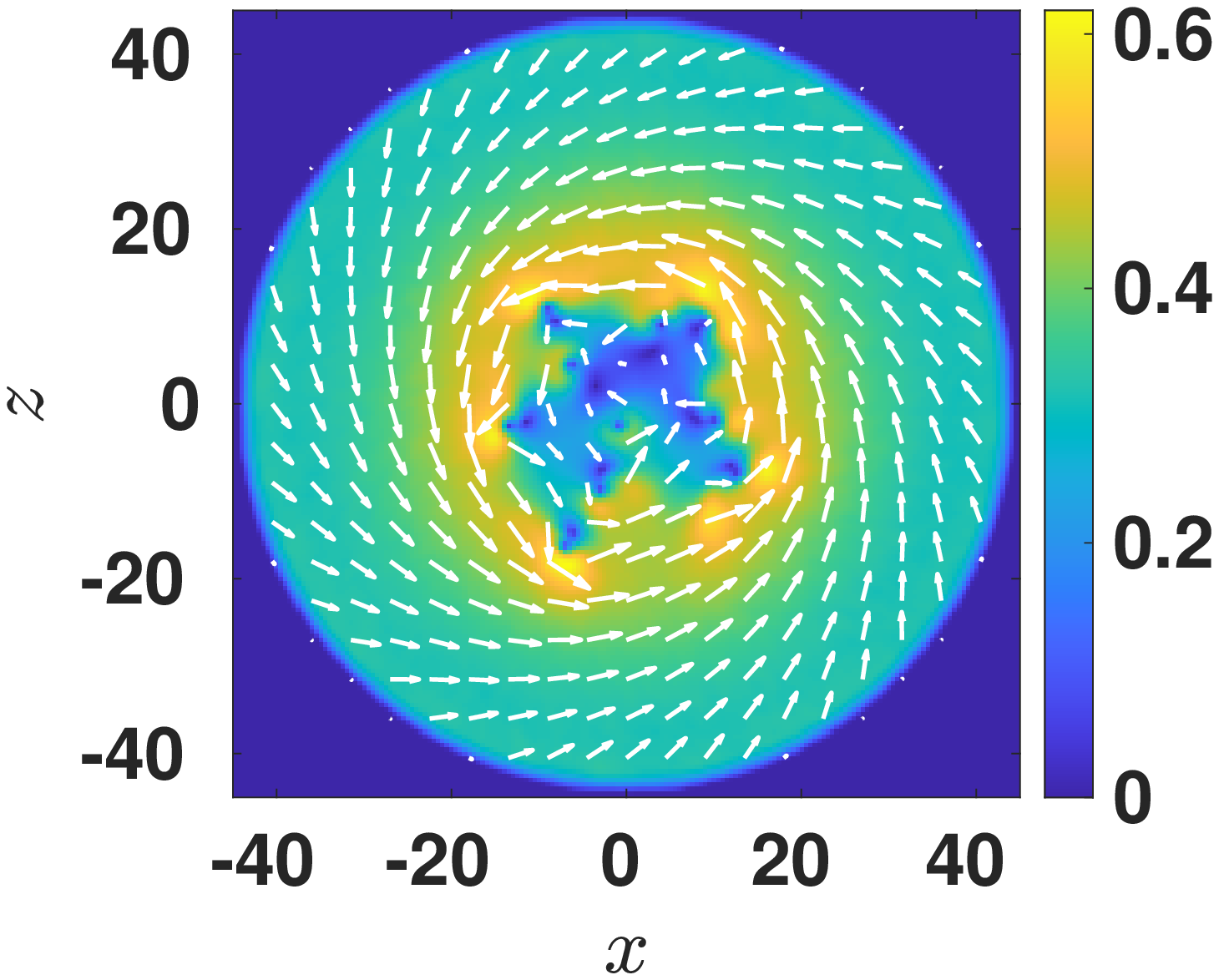}
\caption{
Top (a): Steady drain flow pattern in the presence of vortex lines
plotted on the $xz$ plane for $y=0$. The arrows indicate the 
direction of the current and the colours represent the current's magnitude.
For clarity, only the flow in the cylinder is plotted,
ignoring the drain hole and the injection annulus.
Bottom (b): Steady drain flow pattern 
plotted on the $xy$ plane at $z=(z_3+z_4)/2$. The parameters are
as in Fig.~\ref{fig:3}.
%$z_1=-107$, $z_2=109$, $z_3=z_1=-107$, $z_4=z_1+7=-100$,
%$r_{ad}=42$, $r_{in}=44.5$, $r_{sink}=5$, $r'_{sink}=30$, and
%$V_{sink}^0=0.1$
}
\label{fig:8}
\end{figure}

An important consequence of the twisted nature of vortex lines is that 
it induces a downwards axial superflow into the drain hole, 
stronger than the axial
superflow caused by the drain hole without vortex lines.
Fig.~\ref{fig:8} (a) and (b) shows the current on the $xz$ and $xy$ 
planes respectively.
The presence of the strong axial flow towards the drain hole 
which is induced by the 
vortices is evident, especially when compared to the axial flow when vortices
are absent: compare Fig.~\ref{fig:3} (a) and Fig.~\ref{fig:8} (a). 
The azimuthal flow induced by vortices can be observed
in Fig.~(\ref{fig:8}) (b).
As the twist is transported along the $z$ direction 
by Kelvin waves, it increases the downward
axial flow in the upper part of the bundle. Probably this is the
mechanism which is responsible for the formation of a central drainpipe
funnel when the helium has a top free surface, as observed by Yano and
collaborators in their experiments \cite{Yano}. Further
numerical simulations with $N_v=25$ vortices confirm the scenario which
we have described.

\section{\label{sec:discussion}Discussion and conclusion}
%\label{sec:discussion}

We have used the Gross-Pitaevskii equation to model
the superfluid drain vortex in its simplest form: at $T=0$
(i.e. without any normal fluid and associated viscous 
effects) and without complications arising from a free surface.
We have found that the superfluid drain vortex consists
of a moderately disordered
twisted bundle of quantised vortex lines. The twist is generated
by the radial inflow into the drain, which is stronger near the drain
which brings the vortex lines closer to each other. We also found that
the twist of the vortex lines
induces a strong central axial flow into the drain;
in the presence of a free surface at the top, this
axial flow is probably the origin of the funnel or drainpipe which has 
been observed \cite{Yano}.

The drain flow twisted bundle is similar to other twisted vortex states which
have been observed in superconductors \cite{Kamien}, and, more relevantly,
in rotating $^3$He \cite{Eltsov} and large-scale $^4$He vortex rings \cite
{Wacks,Galantucci,Svancara}. 

Finally, by combining order (the strong axial polarization) with
a controlled amount of disorder (induced by the radial drain flow), our
results show that this
bathtub vortex flow is a promising configuration
to develop theoretical tools 
based on the HVBK equations and Vinen equation \cite{Lipniacki2006,Jou2011}
to model problems of rotating quantum turbulence \cite{Walmsley2012}, 
ranging from liquid helium counterflow
\cite{Swanson1983,Tsubota2003} to neutron star glitches 
\cite{Mongiovi2017,Haskell2020} to atomic gases \cite{Hossain2022}.

We are grateful to Sam Patrick and Silke Weinfurtner for discussions
and acknowledge the financial support of UKRI grant  
{\it Quantum simulators for fundamental physics} (ST/T006900/1).
LG acknowledges the support of Istituto Nazionale di Alta Matematica (INdAM).

%\begin{appendices}
%\section{Section title}\label{secA1}
%An appendix contains supplementary information that is not an essential 
%\end{appendices}

%\bibliography{sn-bibliography}% common bib file

\newpage
\begin{thebibliography}{10}
\bibitem{Vinen1961}
W.F. Vinen, The detection of single quanta of circulation in liquid helium~II,
Proc. Roy. Soc. London A {\bf 260}, 218 (1961).

\bibitem{Andersen}
A. Andersen, T. Bohr, B. Stenum, J. Juul Rasmussen, and B. Lautrup,
Anatomy of a bathtub vortex,
Phys. Rev. Letters. {\bf 91}, 104502 (2003).

\bibitem{Bohling}
L. Bohling, A. Andersen, and D. Fabre,
Structure of a steady drain-hole vortex in a viscous fluid,
J. Fluid Mech. {\bf 656}, 177 (2010).

\bibitem{Yano}
H. Yano, K. Ohyama, K. Obara, and O. Ishikawa,
Observation of the spiral flow and vortex induced by  suction
pump in superfluid $^4$He,
J. Phys. Conf. Series {\bf 969}, 012002 (2018).

\bibitem{Matsumura}
I. Matsumura, K. Ohyama, K. Sato, K. Obara, H. Yano, and O.
Ishikawa,
Observation of second sound attenuation across a
superfluid suction vortex,
J. Low Temp. Physics {\bf 196}, 204 (2019).

\bibitem{Obara}
K. Obara, I. Matsumura, N. Tajima, K. Ohyama, H. Yano, and O. Ishikawa,
Vortex line density of superfluid suction vortex,
Phys. Rev. Fluids {\bf 6}, 064802 (2021).

\bibitem{Inui}
S. Inui, T. Nakagawa, and M. Tsubota,
Bathtub vortex in superfluid $^4$He,
Phys. Rev. E {\bf 102}, 224511 (2020).

\bibitem{Schwarz}
K.W. Schwarz,
Three-dimensional vortex dynamics in superfluid $^4$He:
homogeneous superfluid turbulence,
Phys. Rev. B {\bf 38}, 2398 (1988).

\bibitem{Primer}
C.F. Barenghi and N.G. Parker, {\it A primer on quantum fluids},
Springer (2016).

\bibitem{StringariPitaevskii}
L. Pitaevskii and S. Stringari, {\it Bose-Einstein Condensation},
Clarendon Press, Oxford (2003).
%\bibitem{Barenghi}
%C.F. Barenghi, L. Skrbek, and K.R. Sreenivasan,
%Introduction to quantum turbulence,
%Proc. Nat. Acad. Sci. USA, {\bf 111} (Suppl. 1), 4647 (2014).

\bibitem{Feynman}
R.P. Feynman,
Application of quantum mechanics to liquid helium,
in {\it Progress in Low Temperature Physics},
vol. 1, ed. C.J. Gorter, North Holland, Amsterdam (1955).

\bibitem{Kamien}
R.D. Kamien,
Force-free configurations of vortices in high-temperature superconductors 
near the melting transition,
Phys. Rev. B {\bf 58}, 8218 (1998).

\bibitem{Eltsov}
V.B. Eltsov, A.P. Finne, R. H\"{a}nninen, J. Kopu, M. Krusius, M. Tsubota,
and E.V. Thuneberg, Twisted vortex state,
Phys. Rev. Lett. {\bf 96}, 215302 (2006).

\bibitem{Wacks}
D.H. Wacks, A.W. Baggaley and C.F. Barenghi,
Coherent laminar and turbulent motion of toroidal vortex bundles
Phys. Fluids {\bf 26} , 027102 (2014).

\bibitem{Galantucci}
L. Galantucci, G. Krstulovic, C.F. Barenghi,
Friction-enhanced lifetime of bundled quantum vortices,
arXiv:2107.07768 (2021).

\bibitem{Svancara}
P. Svancara, M. Pavelka and M. La Mantia,
An experimental study of turbulent vortex rings
in superfluid $^4$He
J. Fluid Mech. {\bf 889}, A24 (2020).

\bibitem{Lipniacki2006}
T. Lipniacki,
Dynamics of superfluid $^4$He: two-scale approach,
European J. Mech.  B/Fluids {\bf 25} 435, (2006).

\bibitem{Jou2011}
D. Jou, M.S. Mongiov\`i and M. Sciacca,
Hydrodynamic equations of anisotropic, polarized and inhomogeneous 
superfluid vortex tangles,
Physica D {\bf 240}, 249 (2011).

\bibitem{Walmsley2012}
P. M. Walmsley and A. I. Golov,
Rotating quantum turbulence in superfluid $^4$He in the $T=0$ limit,
Phys. Rev. B {\bf 86}, 060518 (2012)


\bibitem{Swanson1983}
C.E. Swanson, C.F. Barenghi and R.J. Donnelly,
Rotation of a tangle of quantized vortices in HeII,
Phys. Rev. Lett. {\bf 50}, 190 (1983).

\bibitem{Tsubota2003}
M. Tsubota, T. Araki and C.F. Barenghi,
Rotating superfluid turbulence,
Phys. Rev. Lett. {\bf 90}, 205301 (2003).

\bibitem{Mongiovi2017}
M.S. Mongiov\`i, F.G. Russo, and M. Sciacca,
A mathematical description of glitches in neutron stars
M.N.R.A.S. {\bf 469} 2141, (2017).

\bibitem{Haskell2020}
B. Haskell, D. Antonopoulou, and C.F. Barenghi,
Turbulent, pinned superfluids in neutron stars and pulsar glitch recoveries,
M.N.R.A.S. {\bf 499}, 161 (2020).

\bibitem{Hossain2022}
K. Hossain, K. Kobuszewski, M. McNeil Forbes, P. Magierski, K. Sekizawa, 
and G. Wlazłowski,
Rotating quantum turbulence in the unitary Fermi gas,
Phys. Rev. A {\bf 105}, 013304 (2022).

\end{thebibliography}
%% if required, the content of .bbl file can be 
%included here once bbl is generated
%%\input sn-article.bbl
%% Default %%
%%\input sn-sample-bib.tex%

\end{document}